\documentclass[11pt,a4paper]{article}

\usepackage{jheppub}
\usepackage{basket}
\usepackage{mathShrtcts}
\usepackage{multirow}

\title{Ambitwistor Strings and the Scattering Equations on AdS$_3\times$S$^3$}

\author{Kai Roehrig and David Skinner}
\affiliation{Department of Applied Mathematics \& Theoretical Physics \\
        University of Cambridge \\
        Wilberforce Road \\
        Cambridge CB3 0WA, United Kingdom}

\date{\today}

\begin{document}
\abstract{
We construct an ambitwistor string that describes Type II supergravity on AdS$_3\times$S$^3$ with pure NS flux. The background Einstein equations ensure that the model is anomaly free. The spectrum consists of supergravity fluctuations around this background, with no higher string states. This theory transforms the problem of computing $n$-point tree-level amplitudes on AdS$_3$ into that of understanding an $\mathfrak{sl}_2$ Gaudin integrable system, whose representations are determined by the dual boundary operators and whose spectral parameters correspond to the worldsheet insertion points. The scattering equations take a similar form to flat space, with $n(n-3)/2$ parameters $\tau_{ij}$ parametrizing the eigenvalues of the Gaudin model.}

 \maketitle

 \section{Introduction}

Recent years have seen impressive developments in understanding boundary correlators on  anti de Sitter space~\cite{Rastelli:2016nze,Rastelli:2017udc,Arutyunov:2018tvn,Alday:2017vkk,Bissi:2020wtv}. Powerful techniques include use of the Mellin representation of AdS amplitudes~\cite{Mack:2009mi,Paulos:2011ie,Fitzpatrick:2011ia,Goncalves:2014rfa}, exploiting hidden, higher dimensionsal conformal symmetries~\cite{Caron-Huot:2018kta}, and bootstrapping the large-$N$ correlators using constraints from the boundary OPE~\cite{Caron-Huot:2017vep,Aprile:2017bgs,Aharony:2016dwx}. As with on-shell methods in flat space, these techniques typically eschew individual Feynman / Witten diagrams,  focussing instead on properties of the correlator as a whole. See~\cite{Rastelli:2019gtj,Giusto:2019pxc} for applications of these ideas to four-point functions in AdS$_3\times$S$^3$.

\medskip

The aim of this paper is to initiate a new approach to $n$-particle boundary correlators in AdS$_3\times$S$^3$, using ambitwistor strings~\cite{Mason:2013sva}. These are perturbative descriptions of field theories, including supergravity, that are nonetheless based on mappings from a world{\it sheet}, rather than a worldline. They therefore enjoy the usual property of string theory that, at any given order in perturbation theory, all $n$-point amplitudes arise from a single worldsheet topology, capturing the sum of all Feynman diagrams.

The reason such strings can describe field theories is because their worldsheet theories are purely chiral. The basic action is
\begin{equation}
\label{eqn:flat-action}
	S = \int_\Sigma P_\mu\,\partialb X^\mu + \tilde {e}\,H\,,
\end{equation}
where $(P_\mu,X^\nu)$ define a map to the (complexified) cotangent bundle of space-time while $H\equiv P^2/2$. The Lagrange multiplier $\tilde{e}$ thus enforces the constraint that $P_\mu$ is null. Accounting for this constraint and the corresponding gauge redundancy, the action~\eqref{eqn:flat-action} describes a map to the space of null geodesics, known as ambitwistor space. After quantization, plane wave vertex operators $\sim {\rm e}^{{\rm i}k\cdot X}$ are BRST invariant if and only if $k$ is null, and the ambitwistor string spectrum consists entirely of massless states. At genus zero, inserting $n$ such  vertex operators $\prod_i {\rm e}^{{\rm i}k_i\cdot X(z_i)}$ and integrating out the field $X$, one learns  that 
\[
	P_\mu(z) ~= ~\sum_i \frac{k_{i\mu}}{z-z_i} \qquad\text{and hence}\qquad
	H(z) ~=~ \frac{1}{2}\,{\sum_{i, j}} \frac{s_{ij}}{(z-z_i)(z-z_j)}\,,
\]
where $s_{ij} = k_i\cdot k_j$ is a Mandelstam invariant. The diagonal terms in the double sum vanish since $k_i^2=0$. In the presence of such insertions the gauge field $\tilde{e}$ has $n-3$ moduli. Performing the path integral over these 
moduli~\cite{Adamo:2013tsa,Ohmori:2015sha} imposes the scattering 
equations~\cite{Gross:1987kza,Cachazo:2013hca} 
\begin{subequations}
\begin{equation}
\label{eqn:flat-space-scattering}
	H_i = 0	
\end{equation}
for any $n-3$ choices of $i\in\{1,2,\ldots,n\}$, where
\begin{equation}
	H_i \equiv {\rm Res}_{z_i} \,H(z) ~ = ~ \sum_{j\neq i} \frac{s_{ij}}{z_i-z_j} ~.
\end{equation}
\end{subequations}
$H(z)$ is a meromorphic quadratic differential, so once $n-3$ of its residues vanish,  $H(z)$ itself vanishes identically throughout $\Sigma$. This ensures the validity of the ambitwistor interpretation even in the presence of vertex operators. The $n-3$ scattering equations~\eqref{eqn:flat-space-scattering} are M{\"o}bius invariant, and localize the integral over the worldsheet insertion points.

This basic story has been extended in many ways. A `type II' RNS ambitwistor string describes type II supergravity in $d=10$, with the CHY integrand~\cite{Cachazo:2013hca,Cachazo:2013iea} for $n$-point amplitudes arising from the worldsheet correlator. At higher genus,  the $n+3g-3$ worldsheet moduli are again localized by genus $g$ analogues of the scattering equations~\cite{Adamo:2013tsa,Geyer:2015jch,Roehrig:2017gbt}. The resulting worldsheet correlators have been shown to correctly reproduce gravitational scattering amplitudes through 2-loops~\cite{Geyer:2016wjx,Geyer:2018xwu}. Other ambitwistor strings have been constructed~\cite{Ohmori:2015sha,Casali:2015vta}, describing amplitudes in range of target space field theories including Yang-Mills, bi-adjoint scalars and certain NLSMs with amplitudes given by the CHY formul\ae\ \cite{Cachazo:2014xea}. Importantly for this paper, ambitwistor strings with generic curved backgrounds have been constructed in~\cite{Adamo:2014wea,Adamo:2018ege}, while 3-point amplitudes were computed in~\cite{Adamo:2017sze} in the specific case of a Ricci-flat plane-wave background. Pure spinor versions of the ambitwistor string have been studied in~\cite{Berkovits:2013xba,Chandia:2015sfa,Adamo:2015hoa,Azevedo:2016zod}.

\medskip

In this paper, we study the ambitwistor string on an AdS$_3\times S^3$ target. There is again a constraint $H(z)=0$, where $H$ now generalizes $P^2$ to the non-Abelian case. {\it A priori}, $H$ is an operator in the worldsheet CFT. However, when the vertex operators represent bulk-to-boundary propagators, it can be represented by an operator acting only on the external data {\it i.e.} the corresponding {\it boundary} locations. Doing so, we find that $H(z)$ may be interpreted as the Hamiltonian of the Gaudin model, a well-known quantum integrable spin chain (see {\it e.g.}~\cite{Gaudin:1976sv,GaudinBook,Sklyanin:1987ih,Sklyanin:1997zz,Feigin:1994in}). In our case, the `spins' are determined by the dual boundary operators, while the worldsheet coordinate $z$ plays the role of the spectral parameter for the Gaudin model.

Despite the non-Abelian nature, integrability ensures that the residues of $H(z)$ are mutually commuting operators, $[H_i,H_j]=0$. Remarkably, we find that the vertex insertion points on the AdS$_3\times S^3$ ambitwistor string are again localized to solutions of scattering equations that take the same form
\begin{equation}
	\sum_{j\neq i} \frac{\tau_{ij}}{z_i-z_j} = 0 
\end{equation}
as in flat space. Here the $n(n-3)/2$ parameters $\tau_{ij}$ label the eigenvalues of the Gaudin Hamiltonians and are closely related to Mellin parameters for the AdS amplitude. The problem of computing $n$-point tree-level amplitudes, {\it i.e.} the sum over all $n$-point tree-level Witten diagrams in supergravity, thus reduces to expanding the result of the worldsheet correlator in terms of a suitable basis of mutual eigenfunctions of the $H_i$. 

\medskip

Here is the structure of this paper. We begin in sections~\ref{sec:classical} \&~\ref{sec:quantization} by constructing the worldsheet theory at the classical and quantum levels. We treat AdS$_3\times S^3$ as a group manifold, and explore the worldsheet algebra that imposes the ambitwistor constraints, generalizing the flat space currents of~\cite{Mason:2013sva}. The type II theory is shown to be anomaly free as a consequence of the Einstein equations, where the target geometry is supported by pure NS flux. In section~\ref{sec:vertex} we study the vertex operators, showing that the spectrum of our theory describes supergravity states travelling on the AdS$_3\times$S$^3$ background.  As a byproduct, in section~\ref{sec:YM-scalar} we construct a heterotic ambitwistor string that we conjecture describes Yang-Mills-Chern-Simons theory on AdS$_3\times$S$^3$, coupled to a higher-derivative gravity. We carry out the worldsheet CFT path integral in section~\ref{sec:correlators}, obtaining the AdS analogues of the flat space CHY Pfaffians. We do this for general gluon polarizations, but limit ourselves in this paper to the simplest case of all-holomorphic graviton polarizations. We also conjecture a formula for the double leading-trace contribution to the $n$-point bi-adjoint scalar amplitude. We turn to understanding the scattering equations themselves in section~\ref{sec:scattering-equations}. We explain the connection to the Gaudin model, and see evidence that the parameters $\tau_{ij}$ are closely related to Mellin parameters.  While we have not found the form of the required Gaudin eigenstates, and so do not have complete description of the amplitudes, we perform  some basic consistency checks in the simplest case of $n=4$. The paper concludes in section~\ref{sec:conclusions} with a discussion of various open questions and possible extensions of this work. Some useful facts about the geometry of group manifolds are collected in appendix~\ref{app:group-manifolds}.

\bigskip

{\it Note added:} While this paper was in preparation, we learned of beautiful, closely related work~\cite{Eberhardt:2020draft} by Eberhardt, Komatsu and Mizera. We thank these authors for sharing a preliminary draft of their paper and for agreeing to coordinate release.

\section{The classical worldsheet theory}
\label{sec:classical}

We begin by describing classical aspects of our model. The bosonic fields of the model are a map $g:\Sigma\to \cal {G}$ from the worldsheet Riemann surface to a Lie group $\cal {G}$, together with a field $j\in\Omega^0(\Sigma,K_\Sigma)\otimes \mathfrak{g}^*$ that is a (1,0)-form on $\Sigma$, taking values in the dual of the Lie algebra $\mathfrak{g}$ of $\cal G$. In terms of a local holomorphic coordinate $z$ on $\Sigma$ and a basis $\hat{t}^a$ of $\mathfrak{g}^*$, we often write $j =j_a(z,\bar z)\,\hat{t}^a = dz\, j_{za}(z,\bar z) \,\hat{t}^a$. We emphasize that, unlike the usual 2d principal chiral model or WZW model, here $j$ and $g$ are independent fields. We also pick a constant, non-degenerate metric $\mathrm{m}(\,\cdot\,,\,\cdot\,)$ on $\cal{G}$ that is bi-invariant, $\mathrm{m}([a,b],c) = \mathrm{m}(b,[c,a])$ for any $a,b,c\in \mathfrak{g}$. On a simple Lie group, such an $\mathrm{m}$ would be proportional to the Killing form. However, we will mostly be interested in the case where $\cal G$ is only semi-simple, in which case other, inequivalent choices of $\mathrm{m}$ are possible.  

With these ingredients, the bosonic action is the natural generalization to $\cal G$ of the flat space bosonic action~\eqref{eqn:flat-action}, namely
\begin{equation}\label{eqn:bosonic-action-before-gauging}
	 S[j,g] = \int_\Sigma \, j_a \,(g^{-1} \partialb g)^a  
	 +\frac{1}{2}\tilde{e}\,\mathrm{m}^{-1}(j,j)\,.
\end{equation}
Note that we do not include a Wess-Zumino term. Since $g^{-1} \partialb g\in\Omega^{0,1}(\Sigma,\mathfrak{g})$, the pairing in the kinetic term is canonical, and does not involve a metric on either $\Sigma$ or $\cal G$.  Note that, since the action involves the worldsheet $\partialb$ operator, we must take $\cal G$ to be a complex Lie group. We expect that a real form may be selected by choosing an appropriate integration cycle over which to perform the path integral over $g$. Finally, $\tilde{e}\in \Omega^{0,1}(\Sigma,T_\Sigma)$ is Lagrange multiplier for the quadratic differential 
\begin{equation}
\label{eqn:H-def}
	H_0 = \frac{1}{2}\mathrm{m}^{-1}(j,j)\,,
\end{equation}
generalizing $P^\mu P_\mu/2$ from flat space. This current will play a key role in what follows.

The action is invariant under left- and right translations of $g$, acting as
\begin{equation}
\label{eqn:left-right-translations}	
	g(z,\bar z) \mapsto \tilde{h}(z)\,g(z,\bar z)\, h(z)\,,\qquad\qquad
	j(z,\bar z) \mapsto h^{-1}(z)\,j(z,\bar z)\,h(z)\,,
\end{equation}
where we have used bi-invariance of $\mathrm{m}$. Here, both $h$ and $\tilde h$ must be holomorphic, so we have two copies of a holomorphic Kac-Moody algebra. Note $J$ transforms covariantly under right translations, but is invariant under left translations. This asymmetry arose from choosing to construct the action in terms of $g^{-1}\partialb g$ rather than $\partialb g \,g^{-1}$. Setting $\tilde h=1$ and allowing $h$ to be smooth, the action is no longer invariant, but obeys 
\[
	S[h^{-1}jh,gh] = S[j,g] + \int_\Sigma j\cdot (h^{-1}\bar\partial h)\,,
\]
showing that $j$ is the generator of right translations. Left translations are instead generated by $gjg^{-1}$.

\medskip

Varying the action~\eqref{eqn:bosonic-action-before-gauging} on a compact Riemann surface, we obtain 
 \begin{equation}
\delta S = \int_\Sigma  \delta j\,\cdot\,\left( g^{-1} \partialb g + \tilde{e} \,\mathrm{m}^{-1}(\,\cdot\,,j)\right)-\left(\partialb j +[\,g^{-1}\partialb g,j\,]\right) \cdot ( g^{-1} \delta g ) + \delta\tilde e\,H_0\,.
 \end{equation} 
Thus, when $\tilde e=0$, the equations of motion simply say that
\begin{equation}
\label{eqn:worldsheet-equations-of-motion}
\begin{gathered}
   \partialb g = 0\,,\qquad\text{and}\qquad
   \bar{D}j=\partialb j + [\,g^{-1}\partialb g,j\,]=0\,.
\end{gathered}
\end{equation}
If the worldsheet is a compact Riemann sphere and no sources are present, these equations are uniquely solved by $g(z)=g_0\in \cal G$ a constant, and $j=0$. If we allow the Riemann surface to have a boundary, varying the action leads to a boundary term $\oint_{\partial\Sigma}j\cdot g^{-1}\delta g$. Further varying this boundary term yields the (pre-)symplectic form on the space of classical solutions as 
\begin{equation}
\label{eqn:symplectic-form}
\begin{gathered}
\Omega =  \oint_{S^1}  \delta j \cdot (g^{-1} \delta g ) -\frac{1}{2}j\cdot[\,g^{-1} \delta g\, ,g^{-1} \delta g\,] \,.
\end{gathered}
\end{equation}
The Poisson brackets associated to $\Omega$ are
\begin{equation}
\label{eqn:bosonic-Poisson}
\begin{aligned}
\{ j_a(\sigma) , g(\sigma') \}  &=   \delta(\sigma-\sigma')\, g(\sigma)  t_a\\
\{ j_a(\sigma) , j_b(\sigma') \}   &=  \delta(\sigma-\sigma')\,\f abc \, j_c(\sigma)
\end{aligned}
\end{equation}
Notice in particular  that, unlike a WZW model, we have $\{g(\sigma),g(\sigma')\}=0$ in the ambitwistor string. These Poisson brackets confirm that $j$ is the Kac-Moody current generating right translations on $\cal G$.

Given a smooth (1,0) vector $v$ on $\Sigma$ we can construct the charge $\oint v H_0$ associated to the current $H_0$. The Poisson brackets above show that, classically, this charge generates the transformations
\begin{subequations}
\begin{equation}
\label{eqn:H-transformations}
	\delta g = v \mathrm{m}^{ab}(g t_a)\,j_b\,,
	\qquad\qquad
	\delta j = 0
\end{equation}
on the matter fields, while $\tilde{e}$ plays the role of a gauge field and varies as \begin{equation}
	\delta \tilde e = -\partialb v
\end{equation}
\end{subequations}
Thus, after imposing the constraint $H_0=0$ and quotienting by the associated gauge transformations, we should consider points on $\cal{G}$ that differ by right translation along a cotangent vector $j$ that is $\mathrm{m}$-null to be equivalent. This shows that the (bosonic) target space of the model is really the space of $\mathrm{m}$-null geodesics in the complex Lie group $\cal G$, which gives the ambitwistor string its name.

\subsection{Type II ambitwistor strings}
\label{sec:worldsheet-fermions}

Following the flat space ambitwistor string, to describe supergravity in the target space, we consider a `Type II' ambitwistor string that also includes two sets of worldsheet fermions. We let $\psi,\, \psi' \,\in \Omega^0(\Sigma ,  K_\Sigma^{\sfrac{1}{2}} \otimes \mathfrak{g})$ be $\mathfrak{g}$-valued fermionic worldsheet spinors, with $\psi (z,\bar z) = \psi_z^a (z,\bar z) \,  \sqrt{\diff z}  \, t_a$ in local coordinates. We stress that, as in the ambitwistor string on flat space, both sets of fermions are holomorphic / left-moving.

Including these fermions, the most obvious kinetic action  is
 \begin{equation}
 \label{eqn:covariant-fermion-action}
S_{\rm kin}[j,g,\psi,\psit] = \int_\Sigma j\cdot(g^{-1}\partialb g) 
+ \frac{1}{2} \, \mathrm{m} (\psi ,\bar{D}\psi)
+\frac{1}{2} \, \mathrm{m} (\psit,\bar{D}\psit)\,,
 \end{equation}
using the metric\footnote{Comparison with the flat space ambitwistor string guides us to use the same metric $\mathrm{m}$ for the fermion kinetic term as was used in  $H$.}  $\mathrm{m}$ and covariant derivative $\bar{D}\psi = \partialb  \psi  + [ \,g^{-1} \partialb g , \, \psi \, ]$, with a similar covariant derivative for $\psit$. Under the left- \& right- translations $g\mapsto \tilde h g h$ of~\eqref{eqn:left-right-translations}, the fermions transform  as
\[
	\psi(z,\bar z) \to h^{-1}(z)\, \psi(z,\bar z)\,h(z)\qquad\qquad
	\psit(z,\bar z) \to h^{-1}(z)\, \psit(z,\bar z)\,h(z)
\]
so both $\psi$ and $\psit$ are in the adjoint under right translations, but invariant under left translations. For some purposes, it is more convenient to have one set of the fermions be invariant under right translations, transforming instead under left-translations. To this end, we introduce a modified current
\begin{equation}
\label{eqn:new-J}
	J_a = j_a + \frac{1}{2} \mathrm{f}_{abc}\psit^b\psit^c
\end{equation}
where $\mathrm{f}_{abc}=\mathrm{m}_{ad}\f bcd$ are totally antisymmetric. This aborbs the connection term from $\mathrm{m}(\psit,\bar{D}\psit)$ into the bosonic kinetic term, so that the kinetic action becomes\footnote{Equivalently, we could arrive at the same form of action~\eqref{eqn:noncov-fermion-action} by still using the original current $j$, but introducing $\psi' = g\psit g^{-1}$.}
\begin{equation}
\label{eqn:noncov-fermion-action}
S_{\rm kin}[J,g,\psi,\psit] = \int_\Sigma J\cdot(g^{-1}\partialb g)
+\frac{1}{2} \, \mathrm{m} (\psi ,\bar{D}\psi)
+\frac{1}{2} \, \mathrm{m} (\psit,\partialb\psit)
\end{equation}
in terms of the new current $J$. The Poisson brackets associated to this action are
\begin{equation}
\label{eqn:fermionic-Poisson}
\begin{aligned}
\{J_a(\sigma),g(\sigma')\} &=\delta(\sigma-\sigma') \,g(\sigma)t_a\,,\quad\quad
&\{ J_a(\sigma),J_b(\sigma')\} &= \delta(\sigma-\sigma') \,\f abc J_c(\sigma)\,,\\
\{\psi^a(\sigma),\psi^b(\sigma')\} &= \delta(\sigma-\sigma')\,\mathrm{m}^{ab}\,,
\quad\quad&\{\psit^a(\sigma),\psit^b(\sigma')\} &= \delta(\sigma-\sigma')\,\mathrm{m}^{ab}\\
\{J_a(\sigma),\psi^b(\sigma')\} &= \delta(\sigma-\sigma')\, \f acb \,\psi^c(\sigma)\,,
\quad\quad&\{J_a(\sigma),\psit^b(\sigma')\} &= 0
\end{aligned}
\end{equation}
with all other Poisson brackets vanishing. In particular, the modified current $J$ still generates right translations of $g$, but leaves $\psit$ invariant. At the classical level, introducing the fermions modifies the generator of left translations to 
\[
	g^{-1}\left(j_a + \frac{1}{2}\mathrm{f}_{abc}\psi^a\psi^b
	+\frac{1}{2}\mathrm{f}_{abc}\psit^a\psit^b\right) g 
	= g^{-1}\left(J_a + \frac{1}{2}\mathrm{f}_{abc}\psit^a\psit^b\right) \,.
\]
This leaves $\psi$ invariant but now transforms $\psit$.  

\medskip

As in the flat space ambitwistor string, the type II model is invariant under (global) transformations generated by two fermionic currents as well as the bosonic current extending $H_0$. When written in terms of $J$, the fermionic currents take the  form\footnote{At the classical level, the action is actually invariant under a 1-parameter family of fermionic transformations generated by
\[
	G(\alpha) = J \cdot \psi + \frac{1-3 \, \alpha }{4} \, \mathrm{m}  \! \left( \psi ,  [ \psit , \psit ]  \right)+ \frac{1-\alpha}{4}  \, \mathrm{m}  \! \left( \psi ,  [ \psi , \psi ]  \right)\,,
\]
with a similar freedom in $\tilde G$. This freedom is fixed in the quantum theory such that $\alpha=\sfrac{1}{3}$, which is what we have used in~\eqref{eqn:G-Gt-def}.}
\begin{subequations}
\begin{equation}
\label{eqn:G-Gt-def}
\begin{aligned}
 G  &= J \cdot \psi +\frac{1}{6}  \, \mathrm{m}  \! \left( \psi ,  [ \psi , \psi ]  \right) \\
 \tilde G &= J \cdot \psit -   \frac{1}{6} \, \mathrm{m}  \! \left( \psit ,  [ \psit , \psit ]  \right)
\end{aligned}
 \end{equation}
each of worldsheet holomorphic conformal weight $\sfrac{3}{2}$. The bosonic current requires no modification from the bosonic case beyond the replacement $j\to J$, and we set
\begin{equation}
\label{eqn:H-def}
 H =  \frac{1}{2} \mathrm{m}^{-1}(J,J)\,.
 \end{equation} 
\end{subequations}
The currents $G$, $\tilde G$ generalize the flat space expressions $P_\mu\psi^\mu$, $P_\mu\psit^\mu$ to the non-Abelian case. One can also show that the currents that were used to construct an ambitwistor string on a general curved background in~\cite{Adamo:2014wea,Adamo:2018ege}  reduce to~\eqref{eqn:G-Gt-def}-\eqref{eqn:H-def} in the case that the target is a group manifold. Using the Poisson brackets~\eqref{eqn:fermionic-Poisson} above one finds that, classically, the algebra of these currents closes:
\begin{subequations}
\label{eqn:SL12-gaugealgebra-classical1}
\begin{equation}
\{ G(\sigma) , G(\sigma') \} =  2\,\delta(\sigma-\sigma')\, H(\sigma)  ~, \qquad 
\{\tilde{G}(\sigma) , \tilde{G}(\sigma') \} =2\,\delta(\sigma-\sigma') \, H(\sigma) ~, \end{equation}
and
\begin{equation}   
\label{eqn:SL12-gaugealgebra-classical2}	
\{ G(\sigma) , \tilde G(\sigma') \} = \{ G(\sigma) , H(\sigma') \} =\{ \tilde{G}(\sigma) , H(\sigma')  \} = 0 \,.
 \end{equation}
\end{subequations}
This is the same worldsheet SL$(1|2)$ current algebra as appears in the flat space type II ambitwistor string. 
While these currents bear a close resemblance to the supercurrents and Sugawara stress tensor of a standard supersymmetric WZW model, we emphasise that they are independent of the worldsheet stress tensor of our model. In particular, the SL$(1|2)$ algebra~\eqref{eqn:SL12-gaugealgebra-classical1}-\eqref{eqn:SL12-gaugealgebra-classical2} is not related to worldsheet diffeomorphisms and should not be interpreted as worldsheet supersymmetry. 

\medskip

To construct the type II ambitwistor string, we gauge the currents $(G,\tilde G,H)$ by introducing two fermionic Lagrange multipliers $\chi,\,\tilde \chi \in \Omega^{0,1}(\Sigma, T_\Sigma ^{\sfrac{1}{2}})$ as well as the bosonic multiplier $\tilde e \,\in \Omega^{0,1}(\Sigma,T_\Sigma)$ that transforms like a Beltrami differential. We also allow the worldsheet complex structure to vary by coupling to a further Beltrami differential $\mu$. We take the action of the type II ambitwistor string to be
\begin{equation}
\label{eqn:type-II-action}
	S_{II} = \int_\Sigma J\cdot (g^{-1}\partialb g) + \frac{1}{2}\mathrm{m}(\psi,\bar{D}\psi) + \frac{1}{2}\mathrm{m}(\psit,\partialb \psit) + \mu\,T + \tilde{e} \,H + \chi\,G+\tilde\chi\,\tilde G\,,
\end{equation}
where $T$ is the holomorphic stress tensor
\begin{equation}
\label{eqn:T-def}
	T = J\cdot(g^{-1}\partial g)
	- \frac{1}{2}\mathrm{m}(\psi,D\psi)
	- \frac{1}{2}\mathrm{m}(\psit,\partial\psit)
\end{equation}
of the worldsheet, with $D\psi=\partial\psi + [g^{-1}\partial g,\psi]$. This action is the natural generalisation to a group manifold of the flat space type II ambitwistor action of~\cite{Mason:2013sva,Adamo:2014wea}. As in flat space, it is possible to obtain `heterotic' and purely bosonic ambitwistor strings by replacing one or both sets of worldsheet fermions $\psi$, $\psit$ with some more general worldsheet current algebra, unrelated to the target space. Classically this is straightforward, but on a curved background it can be more subtle  at the quantum level, so we defer its consideration until section~\ref{sec:YM-scalar}.

 \section{Quantization}
\label{sec:quantization}

We now  quantize the theory of the previous section. We being by discussing quantization of the system with $\tilde{e} = \mu = \chi=\tilde\chi=0$, adding the gauge constraints later by the usual BRST  method.

\subsection{OPEs of the fundamental fields}

Upon quantization, the Poisson brackets~\eqref{eqn:bosonic-Poisson}-\eqref{eqn:fermionic-Poisson} become OPEs between the basic fields. These OPEs may be derived in the standard way, but let us highlight a few features that are special to and important in this chiral model. 

Firstly, in contrast to a standard WZW model, the field $g$ here has no non-trivial OPEs other than with $J$. This is because, in the absence of any insertions of $J$, we can integrate out $J$ and localize the remaining correlator onto classical configurations where $g(z)$ is holomorphic throughout $\Sigma$. Even for a non-compact Riemann surface, this shows that such correlators are regular everywhere as a function of the locations of $g$. Thus the OPEs
\begin{subequations}
\begin{equation}
\label{eqn:g-OPEs}
 g(z) \, g(w)   \sim   0  ~,   \qquad g (z) \, \psi^a (w) \sim   0~, \qquad    g (z) \, \psit^a (w)  \sim 0 
 \end{equation}
must all be regular. Regularity of the $gg$ OPE is a characteristic feature of ambitwistor string models, and makes them dramatically easier to work with than a standard WZW model or the full AdS$_3\times S^3$ string.

\medskip

As usual, OPEs involving $J$  may be derived by considering Ward identities for the right translations that $J$ generates. We find that the $Jg$,  $J\psi$ and $J\psit$ OPEs are
\begin{equation}\label{eqn:J-OPEs}
 J_a(z) \, g(w)  ~ \sim ~ \frac{g(w)t_a}{z-w} \,,\qquad
 J_a (z) \, \psi^b (w) ~ \sim ~ -\frac{\f acb   \, \psi^c(w)}{z-w}  \,,\qquad
 J_a(z)\,\psit^b(w) ~\sim ~0
 \end{equation}
 which, like their classical Poisson brackets, just represent the transformation properties of these fields. However, the Kac-Moody  algebra for $J$ becomes\footnote{As usual, $\kappa_{ab} = \f acd \f bdc$ is the Killing form on $\cal G$.}
\begin{equation}
\label{eqn:Kac-Moody}
J_a (z) \, J_b (w) ~ \sim ~ -\frac{\sfrac{1}{2}\,\kappa_{ab}}{(z-w)^2} 
~+~\frac{\f abc \, J_c(w)}{z-w}  ~ \,,
\end{equation}
\end{subequations}
acquiring a level $k=-\sfrac{1}{2}$ in the quantum theory. 

As with all quantum anomalies, this level has its origin in the transformation properties of the path integral measure. (Note again that our action does not contain a Wess-Zumino term.) To understand its value, let us temporarily return to the covariant action~\eqref{eqn:covariant-fermion-action} involving the original current $j$. With two sets of covariant fermions, the path integral measure $[Dj\,Dg\,D\psi\,D\psit]$ is invariant under right translations. This can be seen by considering the path integral over non-constant modes of the fields. In particular, let us pick a solution  $j=\psi=\psit=0$ and $g(z)= g_0\in \cal G$ of the classical equations of motion and integrate out fluctuations around this base point. Integrating out fluctuations in the bosons, one obtains a factor of $1/\det'(\partialb_{\mathfrak{g}})$, acting on the tangent space $T_{g_0}{\cal G}\cong\mathfrak{g}$. Under right translation of $g_0$ we have $\partialb_{\mathfrak{g}}~ \mapsto \partialb_{\mathfrak{g}} + \theta$ for some $\theta$, and according to Quillen's construction~\cite{Quillen:1985} the chiral determinant varies as
 \begin{equation}
 \label{eqn:Quillen}
 \delta_\xi \ln  \det \left( \partialb_{\mathfrak{g}}   \right)  ~ \propto     ~ 
 \left\langle \int_\Sigma \f abc \, \left(g^{-1} \partial g \right)^a \, \theta_c^b     \right\rangle~.
 \end{equation} 
This indicates that the path integral measure $[DJ\,Dg]$ is not right invariant by itself. However, we obtain an additional factor of ${\rm Pfaff}(\partialb_{\mathfrak{g}})^2$ in the numerator from each set of fermions $\psi$ and $\psit$. Together, these cancel the anomalous behaviour of the bosons. Hence, with two sets of covariant fermions the path integral measure would be invariant and the Kac-Moody level $k$ would vanish. 

Given that the $jj$ OPE has $k=0$, the level in the $JJ$ OPE is generated by double contractions involving the fermion terms in $J_a = j_a + \mathrm{f}_{abc}\psit^b\psit^c/2$. It is easily checked that this leads to the level $k=-\sfrac{1}{2}$ as in~\eqref{eqn:Kac-Moody}\footnote{Alternatively, had we worked with the original current $j$ but introduced a new fermion $\psi' = g\psit g^{-1}$, the level of the $jj$ OPE would be modified because the transformation $\psit\to\psi'$ introduces as Jacobian in the path integral measure.}. Indeed, exactly this shift in level between covariant and non-covariant fermions is a well-known feature of supersymmetric WZW models, see {\it e.g.}~\cite{Figueroa-OFarrill:1995vqf}. We emphasize that, provided we calculate using the OPEs~\eqref{eqn:g-OPEs}-\eqref{eqn:Kac-Moody}, we can treat $J$ as a fundamental field.

\medskip

The remaining OPEs are those among the fermions, given by
 \begin{equation}
   \psi^a (z) \, \psi^b (w) ~ \sim ~ \frac{\mathrm{m}^{ab}}{z-w}  ~ , \qquad\qquad  
   \psit^a (z) \, \psit^b (w) ~ \sim ~ \frac{\mathrm{m}^{ab}}{z-w}
 \end{equation}
as usual. This is immediate for the $\psit$s, which have a bare kinetic term. For the $\psi$s it may be obtained by computing the fermion propagator, defined as the solution $G^{ab}(w,z)$ to 
\[
   \left(  \delta _b ^a \,  \partialb  +     \f bca    \,  (g^{-1} \partialb g)^c  \right) G^{bd}(w,z  )     = \bar \delta (w-z) \,\mathrm{m} ^{ad}
\]
for arbitrary, fixed $g$. This is just a conjugation of the free fermion propagator and reads
\begin{equation}
G^{ab}(w,z) = \frac{ \sqrt{  \diff w  } \, \sqrt{    \diff z }}{w-z} ~  \mathrm{m}^{-1} \!  \left(  g(w) t^a g^{-1}(w)  , \,  g(z) t^b g^{-1}(z)    \right)\,.
\end{equation}
Expanding for small $w-z$ and using the invariance of the metric yields the standard free fermion propagator and OPE.

 \subsection{Einstein equations from OPEs of the currents}
 \label{sec:Einstein}

With next consider OPEs between the currents $(G,\tilde G,H)$. In the case of the ambitwistor string on a generic curved background~\cite{Adamo:2014wea,Adamo:2018ege}, this required a great deal of care. Firstly, various quantum (derivative) corrections had to be added to the currents of the classical theory in order to ensure they behaved correctly under target space diffeomorphisms at the quantum level. In the special case that the background is a Ricci-flat plane wave, \cite{Adamo:2017sze} showed that these derivative corrections vanish, greatly simplifying the quantum currents. We shall see that there is a similar simplification when the target is a group manifold.

\medskip

We first note that, despite the fact that they are composite operators, provided the metric can be chosen so that $m^{ab}\kappa_{ab}=0$ the currents  
\[
 G  = J \cdot \psi + \frac{1}{6}  \, \mathrm{m}  \! \left( \psi ,  [ \psi , \psi ]  \right)\,,\qquad
 \tilde G = J \cdot \psit -   \frac{1}{6} \, \mathrm{m}  \! \left( \psit ,  [ \psit , \psit ]  \right)
  \quad\text{and}\quad
 H =  \frac{1}{2}\mathrm{m}^{-1}(J,J)
\]
are each free of normal ordering ambiguities.  Put differently, if one defines them by point splitting the various terms, then in each case the limit in which the point splitting is removed both unambiguous and in fact finite without further subtractions. Freedom from normal ordering ambiguity is a strong indication that these currents are correct at the quantum level.

While the operators $(G,\tilde G,H)$ may themselves be well-defined, it is not guaranteed that their OPEs yield the same SL$(1|2)$ algebra we saw for the classical Poisson brackets, because such composite operators may have higher order poles in their OPEs. Any such higher order pole would be indicative of an anomaly, preventing us from gauging this SL$(1|2)$. A careful treatment\footnote{All OPEs in this paper have been checked with the help of the \textsc{mathematica} software package Lambda~\cite{Ekstrand:2010bp}. Obtaining the OPEs~\eqref{eqn:gauge-algebra-OPEs} uniquely fixes all the numerical coefficients in the classical currents.} using the OPEs of the previous section shows that
 \begin{equation}
 \label{eqn:gauge-algebra-OPEs}
 \begin{aligned}
  G(z) \, G(w) &~\sim ~  {}- \frac{1}{3} \, \frac{ \kappa_{ab} \, \mathrm{m}^{ab}}{(z-w)^3}     +  \frac{2 \, H}{z-w} ~, \\
      G(z) \,  \tilde G(w) &~\sim ~  0  ~, \\
  \tilde G(z) \, \tilde G(w)  &~\sim ~  {}- \frac{1}{3} \, \frac{ \kappa_{ab} \, \mathrm{m}^{ab}}{(z-w)^3}     +  \frac{2 \, H}{z-w}  ~.
 \end{aligned}
 \end{equation}
Thus the SL$(1|2)$ worldsheet algebra is anomaly free iff
\begin{equation}
  \kappa_{ab} \, \mathrm{m}^{ab}=0\,.
 \end{equation} 
We have also checked that this condition is sufficient to ensure that  
\begin{equation}
\label{eqn:HH-OPE-trivial}
	G(z)\,H(w)~\sim~\tilde{G}(z)\, H(w)~\sim~H(z)H(w)~\sim~0\,.
\end{equation}
In other words, we can gauge the worldsheet $SL(1|2)$ algebra to form an ambitwistor string iff the group manifold ${\cal G}$ admits a bi-invariant metric $\mathrm{m}$ such that the $\mathrm{m}$-trace of the Killing form vanishes. At least for bosonic groups, this rules out the choice $\mathrm{m}\propto \kappa$, so ${\cal G}$ cannot be simple. 

In this paper, we will be interested in ${\cal G}\cong SL(2,\mathbb{C})\times SL(2,\mathbb{C})$, which contains ${\rm AdS}_3\times S^3$ as a real slice.  Since this group possess two simple factors, it has two linearly independent bi-invariant metrics. These is most easily understood with the help of the isomorphism $\mathfrak{sl}_2\times\mathfrak{sl}_2\cong \mathfrak{so}(4)$ in which every Lie algebra index is exchanged for an antisymmetric pair of auxiliary $\mathfrak{so}(4)$ indices, {\it i.e.} $ t_a \mapsto t_{mn}$ with $t_{mn} =  -t_{nm}$.   A straightforward computation shows that the Killing form on ${\cal G}$ is given by
 \begin{equation}
 \kappa_{ab}  ~ \cong ~ \eta_{mp}  \, \eta_{nq} -  \eta_{mq}  \, \eta_{np} \,,\qquad\text{with}~a \simeq [mn]\,,~ b \simeq [pq]
 \end{equation}
in $\mathfrak{so}(4)$ notation (see {\it e.g.}~\cite{Berkovits:1999im}). 
The inequivalent $\mathfrak{so}(4)$-invariant metric is the Levi-Civita symbol $\varepsilon_{mnpq}$. Any linear combination of the Killing form and Levi-Civita symbol is a candidate for our background metric $\mathrm{m}_{ab}$, but we see that the worldsheet SL$(1|2)$ current algebra is anomaly free iff we choose
  \begin{equation}
  \mathrm{m}_{ab} ~ \cong ~ \varepsilon_{mnpq} ~,  \qquad \text{with} ~ a \simeq [mn] \, ,  ~ b \simeq [pq] 
 \end{equation}
(up to an overall scale). This structure is well known and has been used extensively to study conventional strings on $AdS_3 \times S^3$ and plane waves of limits of $AdS_p \times S^q$ \cite{}.

\medskip

This condition also agrees with the calculations of~\cite{Adamo:2014wea,Adamo:2018ege}, where the worldsheet SL$(1|2)$ current algebra of an ambitwistor string on a generic curved manifold was shown to be anomaly free iff the NS-NS background obeyed the Einstein equations. These are
\begin{subequations}\label{eqn:supergravity-eqn}
 \begin{align}
R_{\mu\nu} - \frac{1}{4} \, H_{\mu\kappa\lambda} \,  H \indices{_\nu ^{\kappa\lambda}} + 2 \,  \nabla_\mu \nabla_\nu \, \Phi = 0 ~,   \label{eqn:supergravity-eqn-graviton}  \\
 \nabla_\kappa H \indices{^\kappa _{\mu\nu}} - 2 \, H \indices{^\kappa _{\mu\nu}}   \, \nabla_\kappa \Phi = 0  ~,    \label{eqn:supergravity-eqn-Bfield} \\
 R + 4 \, \nabla_\mu \nabla^\mu \Phi - 4 \, \nabla_\mu \Phi \,  \nabla^\mu \Phi - \frac{1}{12} H^2 = 0 ~,     \label{eqn:supergravity-eqn-dilaton} 
 \end{align}
\end{subequations}
where $R_{\mu\nu}$ is the Ricci tensor, $H_{\mu\nu\kappa}$ the NS 3-form flux, and $\Phi$ the dilaton. Notice that these equations do not include an explicit cosmological constant, but instead stabilize the geometry using the NS 3-form flux.
On any group manifold it is simple to show that, in the Maurer-Cartan frame, the Riemann tensor 
 \begin{equation}
  R \indices {_{abc}^d} = \frac{1}{4} \, \f abe \, \f ecd
 \end{equation}
and so is determined by the structure constants. This can be seen  {\it e.g.} by evaluating $[ \nabla_a , \nabla_b ]   \, V_c \equiv R \indices{_{abc}^d } \, V_d$. (See appendix~\ref{app:group-manifolds} for a brief review of the relevant geometry of Lie groups.) Consequently, the Ricci tensor is proportional the Killing form, $ R_{ab} = -\tfrac{1}{4} \, \kappa_{ab}$. We emphasize that this holds irrespective of the choice of bi-invariant metric. Given that the AdS$_3 \times S^3$ dilaton is constant in the Maurer-Cartan frame, from \cref{eqn:supergravity-eqn-graviton,eqn:supergravity-eqn-Bfield} we deduce the identification
\begin{equation}
H_{abc} = -  f_{abc}
\end{equation}
for the background NS three-form. Using these identifications, the Einstein equation~\eqref{eqn:supergravity-eqn} and $B$-field equation \eqref{eqn:supergravity-eqn-Bfield} are satisfied automatically, whilst the dilaton equation~\eqref{eqn:supergravity-eqn-dilaton} reduces to
\begin{equation} \label{eqn:algebra-thridorder-pole}
- \frac{1}{3} \, \kappa_{ab} \, \mathrm{m}^{ab} = 0
\end{equation}
 which is precisely the anomalous term in the OPEs \eqref{eqn:gauge-algebra-OPEs}.

\medskip

Let us make a few remarks. Firstly, it is very striking that the current $H$ can take such a simple form $\mathrm{m}^{-1}(J,J)/2$, even when describing supergravity on a curved background. This simple form plays an important role in helping understanding the AdS scattering equations. 
 
Next, notice our anomaly condition fixes $\mathrm{m}$ only upto an overall constant. The Einstein equations~\eqref{eqn:supergravity-eqn-graviton}-\eqref{eqn:supergravity-eqn-dilaton} are independent of this scaling, with the Einstein and $B$-field equations each being homogeneous of degree zero and the dilaton equation homogeneous of degree $-1$ under $\mathrm{m}\to \lambda\,\mathrm{m}$ for $\lambda$ constant. Since $H_{abc}= \mathrm{m}_{ad}\f bdc$ this shows that, unlike in full string theory, neither the $B$-field flux nor the AdS curvature scale are quantized in the ambitwistor string. In particular, the NS flux is unrelated to the Kac-Moody level. This is possible because the ambitwistor string describes pure supergravity, with no $\alpha'$ or higher curvature corrections. Thus, the Newton constant sits in front of the entire target space supergravity action. Classically, there is no preferred Planck scale. Finally, the condition $\mathrm{m}^{ab}\kappa_{ab}=0$ can be recast as the statement that the dual Coxeter number of $\cal G$ must vanish. It would certainly be very interesting to study ambitwistor strings on either cosets or supergroups for which the dual Coxeter number vanishes.

Finally, we note that we can move between the right- and left-Cartan frames by defining fields $J_L = g J g^{-1}$, $\psi_L= g\psi g^{-1}$ and $\psit_L= g\psit g^{-1}$. Since we transform both sets of fermions, changing variables $(J,\psi,\psit)\mapsto (J_L,\psi_L,\psit_L)$ introduces no Jacobian in the path integral measure. In terms of the new fields, the action~\eqref{eqn:noncov-fermion-action} becomes
\begin{equation}
\label{eqn:worldsheet-action-left-frame}
	\int_\Sigma J_L \cdot\partialb g\,g^{-1}+\frac{1}{2}{\mathrm m}(\psi_L,\partialb\psi_L) + \frac{1}{2}{\mathrm m}(\psit_L,\bar{D}\psit_L)\,,
\end{equation}
where $\bar{D}\psit_L = \partialb\psit_L - [\,\partialb g\,g^{-1},\psit_L\,]$. In particular, the roles of the fermions have switched, with $\psi_L$ now having trivial OPE with $J_L$, whilst $J_L$ transforms $\psit_L$. The currents $(G,\tilde G,H)$ take the same form in terms of the left  fields as before, so again the role of $G$ and $\tilde G$ is switched. We see that this is just the same as switching the sign of $\f abc$, which is indeed a parity transform. The type II model  is thus invariant under parity transformations.

\subsection{The Virasoro algebra}
\label{sec:Virasoro}

We now consider the Virasoro algebra generated by the stress tensor~\eqref{eqn:T-def}.  We find that, as expected, the currents~\eqref{eqn:G-Gt-def} transform as primary operators with weights $h=3/2$, whilst $H$ transforms as a primary of weight $h=2$. On the other hand, the $TT$ OPE is
 \begin{equation}
 T(z) T(w) ~\sim ~  \frac{1}{2}\, \frac{3d }{(z-w)^4} + \frac{2T(w)}{(z-w)^2}  + \frac{\partial T(w)}{z-w} 
 \end{equation}
where $d$ is the complex dimension of the target space. There is no antiholomorphic stress tensor or Virasoro algebra in this chiral worldsheet theory. 

\medskip

After BRST quantization, the stress tensor receives a further contribution from the ghost sector. We introduce the usual $(b,c)$ ghosts for worldsheet diffeomorphisms generated by $T$, and a further fermionic ghost system $(\tilde b,\tilde c)$ associated to transformations generated by $H$. We also introduce two pairs of bosonic ghosts $(\beta,\gamma)$ and $(\tilde\beta,\tilde\gamma)$ associated to gauging of the fermionic currents $G$ and $\tilde G$, respectively.  We stress that, as always in the ambitwistor string, all these ghosts are holomorphic. Specifically, we have
\[
	c,\tilde c\in\Pi\Omega^0(\Sigma,T_\Sigma) \,,\quad
	b,\tilde b\in \Pi\Omega^0(\Sigma,K^2_\Sigma)\,,\quad
	\gamma,\tilde\gamma\in\Omega^0(\Sigma,T^{\sfrac{1}{2}}_\Sigma)
	\quad	\text{and}\quad
	\beta,\tilde\beta\in\Omega^0(\Sigma,K_\Sigma^{\sfrac{3}{2}})\,.
\]
The BRST operator is
\begin{equation}
\label{eqn:BRST}
Q = \oint c(T_{\rm m}+T_{\rm gh}) + \tilde c H + \gamma G + \tilde\gamma \tilde G+\frac{1}{2}\tilde{b}\,(\gamma^2+\tilde\gamma^2)
\end{equation}
where the $\tilde b\,(\gamma^2+\tilde \gamma^2)$ term reflects the structure of the SL$(1|2)$ algebra and where 
\begin{equation}
\label{eqn:T-ghost-def}
	T_{\rm gh} = - 2b\,\partial c - \partial b \,c - 2\tilde b\,\partial\tilde c- \partial \tilde b\,\tilde c+\frac{3}{2}\beta\,\partial \gamma + \frac{1}{2}\partial\beta\,\gamma 
	+\frac{3}{2}\tilde\beta\,\partial\tilde\gamma + \frac{1}{2}\partial\tilde\beta\,\tilde\gamma
\end{equation}
is the stress tensor for the ghosts.

Including the ghost contribution, the central charge in the Virasoro algebra is shifted by $-26$ for each pair $(b,c)$ and $(\tilde b,\tilde c)$, and shifted by $+11$ for each pair $(\beta, \gamma)$ and $(\tilde\beta,\tilde\gamma)$. Altogether this yields a central charge
 \begin{equation}
 {\rm c} = 3 d -26 -26 +11+11 = 3 \, (d -10) \,.
 \end{equation}
This is expected since Type II ambitwistor strings describe $d=10$ supergravity. Our $SL(2,\mathbb{C})\times SL(2,\mathbb{C})$ group manifold makes up only six dimensions,  so we must include a further chiral CFT of ${\rm c}=12$. The simplest possibility is a chiral CFT describing an internal Ricci flat four-manifold $M$ (again complexified), so that the target space has a real slice AdS$_3 \times S^3 \times M$. The fields of this internal CFT will modify the form of the currents $G$, $\tilde G$ and $H$. However, in this paper we will focus on states that are independent of the internal CFT, in which case the internal fields simply decouple and can be ignored for the purposes of computations at genus zero.

\medskip

To summarize the results of this section, the type II ambitwistor string has a BRST operator~\eqref{eqn:BRST} which is nilpotent on the target space AdS$_3\times S^3\times M$ with $M$ a Ricci flat four-manifold and where the metric $\mathrm{m}$ on AdS$_3\times S^3$ is $\mathrm{m}_{ab}=\varepsilon_{[mn][pq]}$, and with AdS$_3\times S^3$ supported by NS flux.

 \section{Vertex Operators}
\label{sec:vertex}

In full string theory on AdS$_3\times S^3$, constructing the spectrum of states requires a careful consideration of the integrable representations of affine SL$(2,\mathbb{R})\times SU(2)$, spectral flow, and an understanding of which of these integrable representations obey the Virasoro (and no-ghost) constraints~\cite{Maldacena:2000hw}. In the ambitwistor case, matters are considerably simpler due to the fact that the $gg$ OPE is trivial.   Since the complete background consistency conditions were equivalent to the statement that the metric $\mathrm{m}$ obeys the AdS$_3\times S^3$ supergravity equations with pure NS flux, it will be no surprise that the vertex operators correspond to linearized supergravity states fluctuating around this background. As in the flat space ambitwistor string, the linearized field equations come from requiring the vertex operators have no double pole with the current $H$, rather than from any condition of the Virasoro algebra. Let us now see this explicitly.

\subsection{NS vertex operators}
\label{sec:undescended-vertex}

In this paper, we restrict attention to vertex operators in the NS sector. We first impose a `GSO projection' by gauging the $\mathbb{Z}_2 \times \mathbb{Z}_2$ symmetry given by $\psi \to - \psi$ and $\psit \to - \psit$. (See~\cite{Berkovits:2018jvm} for a consideration of the additional, non-unitary states that enter the ambitwistor string if this projection is not imposed.) Then, similar to the RNS string in flat space, fixed vertex operators take the form
\begin{equation}
\label{eqn:fixed-vertex-operator}
 U = c \, \tilde c \, \delta (\gamma) \, \delta (\tilde \gamma) ~ \psi^a\psit^b\, V_{ab}(g) ~,
 \end{equation}
where $V_{ab}(g)$ is (the pullback to $\Sigma$ of) a second rank tensor on the group manifold ${\cal G}$. These are the fixed vertex operators at `picture number'\footnote{We borrow the same terminology as used in the RNS string, though we again caution the reader that  the worldsheet theory of the type II ambitwistor string is really an SL$(1|2)$ gauge theory. In particular, it does not live on a super Riemann surface. See {\it e.g.}~\cite{Witten:2012bh} for a clear discussion of picture number.} $(-1,-1)$, the picture number on account of the $\delta(\gamma)\,\delta(\tilde\gamma)$ factor. Note that, because the $gg$ OPE is trivial, these vertex operators do not require normal ordering. This makes them significantly easier to construct than in full string theory. In particular, $V_{ab}(g)$ always transforms as a worldsheet scalar, unlike a usual WZW model for which the worldsheet conformal weight depends on the representation. Consequently, \eqref{eqn:fixed-vertex-operator} is itself a worldsheet scalar and so is annihilated by the part of the BRST  operator depending on the stress tensor.

On the other hand, there are non-trivial OPEs between $U$ and the SL$(1|2)$ charges $G, \tilde G, H$. For the vertex operator to be BRST closed, we require that there are no double poles in these OPEs. This imposes the constraints 
 \begin{subequations}
 \begin{gather}
 e^b \, V_{ab} = 0~,\qquad
 e^a \, V_{ab} -  \mathrm{f}_{b}^{ac}  V_{ac}  = 0 ~,  \\
 e^2 \, V_{ab}   -   2 \, \mathrm{f}_{b}^{cd} \, e_c   V_{ad}  + \kappa_{b}^c \, V_{ac} = 0 
 \end{gather}
 \end{subequations}
on the tensor field $V_{ab}$. Here, $e_a \equiv e_a ^\mu \, \partial_\mu$ is a vector field acting on $V_{ab}$ component-wise, so the first two constraints above are first order differential equations, whilst the second line is a second order equation. We can translate these constraints into a more familiar form with the help of the relation \eqref{eqn:spinconnection-structureconstants} between the structure constants and the connection. Doing so yields
 \begin{subequations}\label{eqn:BRSTconstraints}
 \begin{gather}
 \nabla^a \, V_{ab} = \frac{1}{2} \, \mathrm{f}_{b}^{ac} \, V_{ac}   ~,   \qquad    
 \nabla^a \, V_{ba} = \frac{1}{2} \, \mathrm{f}_{b}^{ac} \, V_{ac}  ~,      \label{eqn:BRSTconstraints-firstorder}  \\
    \nabla^2 V_{ab}  +\mathrm{f} _a^{cd} \, \nabla_c V_{db} + \frac{1}{4} \, (\kappa_a^c \, V_{cb} +\kappa_b^c \, V_{ac} ) + \frac{1}{2} \mathrm{f}_{ac}^d \mathrm{f}_b^{ce} V_{de} =0 
   ~,   \label{eqn:BRSTconstraints-secondorder}
 \end{gather}
 \end{subequations}
where $\nabla_a \equiv e_a ^\mu \, \nabla_\mu$ is the Levi-Civita connection of $\mathrm{m}$ in the Maurer-Cartan frame. 

We decompose the tensor $V_{ab} $ as 
 \begin{equation}
 V_{ab} = \delta G_{ab} + \delta B_{ab} + \mathrm{m}_{ab} \, \delta \Phi ~,
 \end{equation}
into its symmetric, anti-symmetric and trace parts. Then the first order equations~\eqref{eqn:BRSTconstraints-firstorder} become the de Donder gauge condition 
on $\delta G$ in the presence of a background NS three-form, together with the transversality condition $\nabla^a\delta B_{ab}=0$.   The second order equation \eqref{eqn:BRSTconstraints-secondorder} decomposes as the linearizations of the supergravity equations \eqref{eqn:supergravity-eqn-graviton}-\eqref{eqn:supergravity-eqn-dilaton} so vertex operators with $(\delta g\,, \delta b,\,\delta\phi)$ describe on-shell fluctuations in the metric, $B$-field and dilaton. There are no further $\mathbb{Z}_2\times\mathbb{Z}_2$-invariant vertex operators in the NS$^2$ sector of the BRST cohomology, unless we allow dependence on the internal CFT describing $M$.

This shows that the exact spectrum of our model encodes on-shell supergravity fluctuations around the AdS$_3 \times$S$^3$ background\footnote{We expect to find the full supergravity spectrum by also including Ramond sectors for each pair of fermions $\psi$, $\psit$.}. As in the ambitwistor string on flat space, there is no Regge trajectory here; the spectrum is pure supergravity, with no higher string states\footnote{Of course, if we allow our vertex operators to depend on the internal CFT describing a Ricci flat four-manifold $M$, we expect towers of Kaluza-Klein states coming from compactifying $d=10$ supergravity on $M$.}.  This reflects the fact that the {\it exact} background consistency conditions were just the Einstein equations, with no $\alpha'$ or higher derivative corrections.

\subsection{Bulk-to-boundary propagators on AdS$_3$}
\label{sec:propagators}

To compute boundary correlation functions in AdS/CFT,  it is natural to take the external states of the AdS scattering process to be bulk-to-boundary propagators. These are $L^2$-normalizable solutions of the linearized equations of motion (without source term) that asymptote to a Dirac $\delta$-function on the boundary of AdS~\cite{Witten:1998qj,Giveon:1998ns,Kutasov:1999xu} and are the natural analogue of plane waves for scattering in AdS. We describe their construction for the AdS$_3$ factor alone, neglecting Kaluza-Klein modes from the $S^3$. See {\it e.g.}~\cite{deBoer:1998gyt,deBoer:1998kjm,Giveon:1998ns,Maldacena:2000hw,Maldacena:2001km,Gaberdiel:2007vu,Cardona:2010qf} for vertex operators describing AdS$_3$ bulk-to-boundary propagators in full string theory.

\medskip

Bulk-boundary propagators are most easily expressed by describing AdS$_3$ using  coordinates $(\gamma, \tilde \gamma, \phi)$, with metric given by\footnote{The coordinates $\gamma$, $\tilde\gamma$ should not be confused with the worldsheet ghosts! Which is meant should be clear from the context.}
 \begin{equation}
 \label{eqn:AdS-metric}
 \diff s^2 ~=  ~ \diff \phi ^2 + e^{2 \phi} \, \diff \gamma \, \diff \tilde \gamma ~.
 \end{equation}
in units where the curvature scale is 1. The identification with $SL(2)$ is provided by
\begin{equation}\label{eqn:ads-sl2-isomorphism}
  g( \phi, \gamma ,\tilde\gamma ) = e^\phi \, \left( \begin{array}{cc}
  \gamma \, \tilde \gamma + e^{-2 \phi} ~&~  \tilde \gamma \\ 
     \gamma ~&~ 1
  \end{array}  \right)  \in SL(2)~.
\end{equation}
This matrix satisfies $\det g = 1$ identically, and the Cartan frame can be computed explicitly via the definition $e^a_\mu = (g^{-1} \partial_\mu g)^a$. In these coordinates, the boundary of AdS$_3$ corresponds to  $\phi \to \infty$. After discarding an infinite overall constant, \cref{eqn:ads-sl2-isomorphism} shows that the boundary is parametrized by matrices of the form
\begin{equation}\label{eqn:ads-boundary-coordinate}
h =  \left( \begin{array}{cc}
  x \,  \tilde  x   ~&~  \bar x \\ 
     x ~&~ 1
  \end{array}  \right)  =  \left( \begin{array}{c}  \tilde x  \\  1 \end{array} \right)  \otimes  \left(  x , \, 1 \right)  ~,
\end{equation}
Since we have stripped away an infinite factor, the overall scale of $h$ is not well defined. The boundary of AdS$_3$ may thus be characterized as
 \begin{equation}
 \partial AdS_3 = \Big\{ h \in \text{Mat}_{2 \times 2}~ | ~ \det h = 0 \, , ~ h \sim r \, h  \Big\} 
\end{equation}
for $r$ a non-zero constant. (This scaling is fixed in~\eqref{eqn:ads-boundary-coordinate} so that $h_{22}=1$.) The determinant condition and scaling redundancy of $h$ means it contains two degrees of freedom. 

Borrowing notation from spinor helicity variables in four dimensions, we can describe a boundary point as
\begin{equation}
\label{eqn:h-spinor-helicity}
	h = |\tilde{\lambda}]\,\langle\lambda|
\end{equation}
where $|\tilde\lambda]$ and  $|\lambda\rangle$ are a pair of two-component projective spinors. Note that, since the overall scale of $h$ is meaningless, $|\lambda\rangle$ and $|\tilde\lambda]$ must be taken to scale independently here (unlike in four dimensional flat space). These rescalings tell us the boundary conformal weights $(\Delta,\bar\Delta)$ of any expression, and they provide powerful constraints on AdS amplitudes. In this way, $|\lambda\rangle$ and $|\tilde\lambda]$ have a dual role as boundary coordinates and representation labels~\cite{Zamolodchikov:1986bd,Teschner:1997ft,Teschner:1997fv,Petersen:1996np,Andreev:1996qn}.  In  particular, the boundary inherits the full SL$(2)\times {\rm SL}(2)$ symmetry of left- and right-translations, which correspond to holomorphic $\times$ antiholomorphic boundary conformal transformations. In our conventions, the translations $g \mapsto g_L\, g\, g_R$ act as
\begin{subequations}
\begin{align}
	 |\lambda\rangle &\mapsto g_R^{-1}|\lambda\rangle\,,
	 &\langle\lambda |&\mapsto \langle\lambda|\,g_R \label{eqn:lambda-trans}\\
	 |\tilde\lambda]&\mapsto g_L|\tilde\lambda]\,,
	 &[\tilde\lambda|&\mapsto [\tilde\lambda|\,g_L^{-1} \label{eqn:lambda-tilde-trans}
\end{align}
\end{subequations}
on the boundary spinor coordinates. \eqref{eqn:lambda-trans}-\eqref{eqn:lambda-tilde-trans} ensure that $\langle \lambda\,\lambda'\rangle$, $[\tilde\lambda\,\tilde\lambda']$, $[\tilde\lambda|g|\lambda\rangle$ and $\langle\lambda|g^{-1}|\tilde\lambda]$   are each invariant under simultaneous global transformations of the bulk and boundary.

In the ambitwistor string, the target space ${\cal G}$ is initially a complex Lie group, so {\it a priori} we must take the coordinates $(\phi,\gamma,\tilde\gamma)$ to be independent complex numbers, treating~\eqref{eqn:AdS-metric} as a holomorphic metric (the complexification of the real AdS$_3$ metric). Likewise, the `boundary' here is really the complexification $\mathbb{CP}^1\times\mathbb{CP}^1$  of the $S^2$ boundary of real AdS$_3$, so $|\lambda\rangle$ and $|\tilde \lambda]$ are independent complex spinors. Taking the chiral path integral over an integration cycle that selects a real form of AdS will also require a reality condition on the boundary data. 

\medskip

Using this notation, the basic bulk-to-boundary propagator for a scalar field dual to an operator in the boundary CFT of conformal weight $(\Delta,\Delta)$ is the well-known expression~\cite{Witten:1998qj,DHoker:1999kzh}
\begin{equation}
\label{eqn:scalar-bulk-to-boundary-propagator}
	\Phi_\Delta(g) ~ = ~ \frac{{\cal C}_\Delta}{[\tilde\lambda|g|\lambda\rangle^\Delta} 
	~ = ~ {\cal C}_\Delta \left( \frac{e^{-\phi } }{e^{-2  \, \phi} + (\gamma -x)  (\tilde \gamma - \tilde x )} \right)^{\Delta}\,,
\end{equation}
where $g\in {\rm SL}(2)$ and where the normalization constant 
\[
	{\cal C}_\Delta = \frac{1}{\pi} \,\frac{\Gamma(\Delta)}{\Gamma(\Delta-1)}
\]
is chosen to ensure $\int_{{\rm AdS}_3} |\Phi_\Delta(g)|^2\,d^3g = 1$. It is well-known that $\Phi_\Delta(g)$ satisfies the linearized equation of motion $(\nabla^2 - m^2_\Delta)\,\Phi_\Delta=0$ for a scalar field of mass $m^2_\Delta = \Delta(\Delta-2)$. 


\medskip 

For our vertex operators we require the bulk-to-boundary propagators of supergravity states. We will focus on the graviton itself. In AdS$_3$ there are two graviton states, dual to the anti-holomorphic and holomorphic stress tensors of the boundary CFT$_2$. To describe them, we introduce the polarization vectors
\begin{equation}
\label{eqn:polarization-spinor-helicity} 
\epsilon_a = \langle\lambda |t_a |\lambda \rangle
\qquad\text{and}\qquad
\bar\epsilon_a(g) = [\tilde\lambda|\,g^{-1}t_a g\,|\tilde\lambda] \,,
\end{equation}
where $t_a$ are our basis of $\mathfrak{sl}_2$. Notice that, via~\eqref{eqn:lambda-trans}-\eqref{eqn:lambda-tilde-trans}, both $\epsilon_a$ and $\bar\epsilon_a(g)$ transform in the adjoint under right-translations and are invariant under left-translations. This is appropriate as our linearized Einstein equations were written in the Cartan frame, and explains why the polarization tensor $\bar\epsilon_a(g)$ must be taken to depend on the bulk coordinate $g$. (Of course, we could have worked with opposite conventions throughout.) To be completely explicit, if we choose the basis
\[
	t_0 = \frac{1}{2}\begin{pmatrix} \,1\, & \,0\, \\ \,0\, & -1 \end{pmatrix}\,,\qquad
	t_+ = \begin{pmatrix} \,0\, & \,1\, \\ \,0\, & \,0\, \end{pmatrix}\,,\qquad
	t_-= \begin{pmatrix} \,0\, & \,0\, \\ \,1\, & \,0\, \end{pmatrix}
\]
and write $\langle\lambda| = (1,-x)$ and $[\tilde\lambda|=(1,-\tilde{x})$ in terms of inhomogeneous coordinates on the boundary sphere, then
\begin{subequations}
\begin{equation}
	\psi(x,z) \equiv \epsilon_a \,\psi^a(z) = \langle\lambda|\psi|\lambda\rangle 
	=\psi^+ + x\,\psi^0  -x^2\psi^-
\end{equation}
while
\begin{equation}
\begin{aligned}
	\psi(\tilde{x},z) &\equiv \bar\epsilon_a (g)\,\psi^a(z) = [\tilde\lambda|\,g^{-1}\psi g\,|\tilde{\lambda}]\\
	&=(g^{-1}\psi g)^++ \tilde{x}\,(g^{-1}\psi g)^0  -\tilde{x}^2(g^{-1}\psi g)^-\,.
\end{aligned}
\end{equation}
\end{subequations}
These expressions may be commonly found in the literature (see {\it e.g.}~\cite{deBoer:1998gyt}). Using the $\mathfrak{sl}_2$ completeness relations, these polarization tensors can be seen to obey
\[
	\epsilon^{(i)}\cdot \epsilon^{(j)}  = \langle i\,j\rangle^2\,,\qquad\qquad
	\bar\epsilon^{(i)}\cdot\bar\epsilon^{(j)} = [i\,j]^2
\]
for any pair of boundary points $|i]\langle i|$ and $|j]\langle j|$. In particular, $\epsilon^{(i)}\cdot\epsilon^{(i)}=\bar\epsilon^{(i)}\cdot\bar\epsilon^{(i)}=0$.

\medskip

Using these  polarization tensors, the two graviton bulk-to-boundary propagators may be expressed as~\cite{deBoer:1998gyt,Kutasov:1999xu} 
\begin{equation}
\label{eqn:graviton-bulk-to-boundary-propagator}
	\delta G_{ab}^+ =\epsilon_a\,\epsilon_b \,\Phi_4(g)
	\qquad\text{and}\qquad
	\delta G_{ab}^- = \bar\epsilon_a(g) \,\bar\epsilon_b(g)\,\Phi_4(g)\,,
\end{equation}
respectively, where $\Phi_4(g) = 1/[\tilde\lambda|g|\lambda\rangle^4$ as in~\eqref{eqn:scalar-bulk-to-boundary-propagator}.  It is readily shown that $\delta g^\pm_{ab}$ each satisfy both the transversality constraints~\eqref{eqn:BRSTconstraints-firstorder} and linearized Einstein equation~\eqref{eqn:BRSTconstraints-secondorder}, again by using~\eqref{eqn:spinconnection-structureconstants} and the $\mathfrak{sl}_2$ completeness relation. These vertex operators depend on the boundary point $|\tilde\lambda]\langle\lambda|$  both through the polarization tensors as well as through $\Phi_4(g)$. This will be important when we come to compute correlation functions. We see that $\delta G^+$ has homogeneity $(0,-4)$ under scalings of $(\lambda,\tilde\lambda)$, while $\delta G^-$ instead has homogeneity $(-4,0)$. We can equivalently consider the combinations $\delta G^+ \,[\tilde\lambda\,\diff\tilde\lambda]^2$ and $\delta G^-\,\langle\lambda\,\diff\lambda\rangle^2$ which are weightless, but now transform as antiholomorphic and holomorphic quadratic differentials on the boundary. This is as expected for the antiholomorphic and holomorphic boundary stress tensors.

To summarize, the picture $(-1,-1)$ vertex operators for the two AdS$_3$ bulk-to-boundary graviton propagators are
\begin{equation}
\label{eqn:graviton-picture-minus-one}
\begin{aligned}
	U^+ &= c\,\tilde c\,\delta(\gamma)\,\delta(\tilde\gamma)\,
	\frac{\epsilon\cdot\psi~\epsilon\cdot\psit}{[\tilde\lambda|g|\lambda\rangle^4}\\
	U^- &=c\,\tilde c\,\delta(\gamma)\,\delta(\tilde\gamma)\,
	\frac{\bar\epsilon(g)\cdot\psi~~\bar\epsilon(g)\cdot\psit}{[\tilde\lambda|g|\lambda\rangle^4}
\end{aligned}
\end{equation}
where $g$ now represents the worldsheet field. We stress again that, unlike in flat space, the vertex operator for $\delta G^-$ depends on $g$ through the polarization tensor $\bar\epsilon_a(g)$ as well as through $\Phi_4(g)$.

\section{Gauge theory on AdS$_3\times$S$^3$}
\label{sec:YM-scalar}

In this section we will briefly consider an AdS version of the `heterotic' ambitwistor string. As in flat space, this is  expected to describe gauge theory coupled to  a higher derivative gravity. We will focus on the gauge sector.

\medskip

The heterotic worldsheet theory may be simply obtained from the type II model by dropping the $\psit$ system, replacing it with some other auxiliary worldsheet current algebra. Dropping the $\psit$s causes no problems. Indeed, $\psit$ decoupled from the other fields of the type II model, both in the non-covariant action~\eqref{eqn:noncov-fermion-action} and OPEs~\eqref{eqn:J-OPEs}. After dropping $\psit$, the $JJ$ OPE still has level $-\sfrac{1}{2}$ because path integral measure for the  $\psit$s, treated as right-invariant, did not affect the anomalous transformation of the measure for the remaining fields. 

Dropping the $\psit$s forces us to drop $\tilde G$, but does not affect $G$ or $H$\footnote{In particular, unlike the internal CFT describing motion on the Ricci flat manifold $M$ present in the type II theory,  the new worldsheet current algebra with which we replace $\psit$ does not play a role in the ambitwistor constraint $H$.}. Since the Kac-Moody level is unchanged, we still have the current OPEs
\begin{equation}
	G(z)\,G(w)\sim \frac{2H(w)}{z-w}\,,\qquad\qquad G(z)\,H(w)\sim H(z)\,H(w)\sim 0
\end{equation}
provided $\mathrm{m}^{ab}\kappa_{ab}=0$, as always. For the same reason, the matter stress tensor
\begin{equation}
	T_{\rm het} = J\cdot (g^{-1}\partial g)- \frac{1}{2}\mathrm{m}(\psi,D\psi) + T_{\rm cur} 
\end{equation}
is still quasi-primary, with central charge ${\rm c}_{\rm cur} + 5d/2$, where c$_{\rm cur}$ is the central charge contributed by the new worldsheet current algebra. 

The BRST operator is
\begin{equation}
\label{eqn:heterotic-BRST}
Q = \oint c(T_{\rm het} + T_{\rm gh}) + \tilde{c}\,H + \gamma \,G + \frac{1}{2}\tilde b\,\gamma^2\,,
\end{equation}
where the ghost stress tensor no longer involves $(\tilde\beta,\tilde\gamma)$. We also gauge the $\mathbb{Z}_2$ that acts non-trivially on $\psi$ and the $(\beta,\gamma)$ ghosts. This model is anomaly-free on an arbitrary ${\cal G}$ provided
\[
	\mathrm{m}^{ab}\kappa_{ab}  = 0\qquad\text{and}\qquad
	{\rm c}_{\rm cur} = 41- \frac{5d}{2}
\]
where ${\rm c}_{\rm cur}$ is the central charge of the current algebra. This central charge condition is exactly the same as for the heterotic ambitwistor string in flat space.  

In the gauge sector, BRST-closed NS vertex operators describing bulk-to-boundary propagators in AdS$_3$ are  
\begin{subequations}
\begin{equation}
\label{eqn:gluon-vertex}
\begin{aligned}
	A^+ &= c \,\tilde c \,\delta(\gamma) \, \tilde\jmath~\Phi_2(g) ~\epsilon\cdot\psi \\
	A^- &=  c \,\tilde c \,\delta(\gamma) \, \tilde\jmath ~\Phi_2(g)~\bar\epsilon(g)\cdot\psi\end{aligned}
\end{equation}
at picture $-1$, where again $\Phi_2(g(z)) =1 / [\tilde\lambda|g(z)|\lambda\rangle^2$. The corresponding picture zero operators are\footnote{We caution the reader that the ordering of the factors in ${\cal A}^-$ is important. In particular, $:\Phi_2(g) \,\bar\epsilon^a(g) J_a: = :\bar\epsilon^a(g)(J_a + \partial g\,g^{1})\,\Phi_2(g):$. No such ambiguity arises in ${\cal A}^+$, the asymmetry being traceable to our original choice of right-Cartan frame.} 
\begin{equation}
\label{eqn:gluon-vertex-picture0}
\begin{aligned}
	{\cal A}^+ &= c\,\tilde c\,\tilde\jmath ~ \Phi_2(g) ~\epsilon\cdot J\\
	{\cal A}^- &= c\,\tilde c\,\tilde\jmath  ~\Phi_2(g) ~\bar\epsilon(g)\cdot J\,.
\end{aligned}
\end{equation}
\end{subequations}
Here, $\tilde{\jmath}$ is a current of the auxiliary $S_{\rm cur}$ carrying worldsheet conformal weight 1.  The combination $A^+\,[\tilde\lambda\,\diff \tilde\lambda]$ is invariant under boundary scalings, and transforms like a $(0,1)$-form on the boundary, whilst the weightless $A^-\,\langle\lambda\,\diff\lambda\rangle$ transforms as a  boundary $(1,0)$-form. As expected, these vertex operators are dual, respectively, to antiholomorphic and holomorphic currents in the boundary CFT$_2$. 

\medskip

Even in the pure gauge sector\footnote{As in flat space~\cite{Mason:2013sva,Berkovits:2018jvm}, there are further $\mathbb{Z}_2$-invariant BRST-closed vertex operators in this model representing gravitational states, with higher derivative interactions.}, it is too hasty to assume this model describes Yang-Mills theory in AdS$_3\times$S$^3$, because of the possibility of Chern-Simons terms. Computing the 2-point function we find
\begin{equation}
\label{eqn:CS-2-pt}
\left\langle\, c \partial c\,\tilde c\, \delta(\gamma)\,\tilde\jmath \,\epsilon_1\cdot\psi\,\Phi_2(g)~c\,\tilde c\partial \tilde c\,\delta(\gamma)\tilde \jmath\, \epsilon_2\cdot\psi\,\Phi_2(g)\,\right\rangle = \frac{1}{[12]^2}\,.
\end{equation}
In AdS$_3$, this 2-point function is not generated by the usual Yang-Mills action, which gives only a contact interaction on the boundary. \eqref{eqn:CS-2-pt} corresponds to a double pole in the boundary current algebra, which indicates the presence of a Chern-Simons term in the bulk.  We therefore propose that the leading-trace sector of this heterotic ambitwistor string describes Yang-Mills Chern-Simons theory in AdS$_3\times$S$^3$.

 \section{Worldsheet correlation functions}
\label{sec:correlators}

We now turn to computing $n$-point correlation functions of vertex operators. These will give the AdS versions of the `CHY integrands' in flat space. Due to the chiral nature of the worldsheet theory, these correlators can be obtained in closed form with relative ease. The path integrals can by performed by repeated application of Ward identities.

\subsection{Correlator of $n$ holomorphic gluons}
\label{sec:n-holomorphic-gravitons}

We begin in the heterotic model, with the correlator 
\begin{equation}
\label{eqn:hol-gauge-corr-undescended}
\mathcal{A}_{++\cdots+}(x_1,\ldots,x_n) = \left\langle A^+_1(z_1)\,A^+_2(z_2)\,\prod_{i=3}^n \,{\cal A}_i^+(z_i) \right\rangle \,.
\end{equation}
of $n$ holomorphic gluons, with two in picture $-1$~\eqref{eqn:gluon-vertex} and $n-2$ in picture 0~\eqref{eqn:gluon-vertex-picture0}. This computation is somewhat simpler than the general case, since all the polarization tensors are independent of the field $g$.  

\medskip

Firstly, the straightforward pieces. As in flat space, at leading trace, the auxiliary worldsheet current algebra simply provides a sum 
\begin{equation}
\label{eqn:Parke-Taylor}
	\sum_{\alpha\in S_n/D_n} {\rm PT}(\alpha)\,,\qquad\text{where}\qquad
	{\rm PT}(\alpha) = \frac{{\rm tr}(T_{\alpha(1)}T_{\alpha(2)}\cdots T_{\alpha(n)})}{z_{\alpha(1)\alpha(2)}z_{\alpha(2)\alpha(3)}\cdots z_{\alpha(n)\alpha(1)}}
\end{equation}
over Parke-Taylor factors with dihedrally inequivalent orderings. Also, as usual, with only two picture $-1$ insertions, $\chi$ may be safely set to zero (its moduli being responsible for the PCOs that changed the remaining operators to picture zero). Integrating over the associated $\beta\gamma$-ghost system gives a factor of $1/z_{12}$.

The worldsheet gauge field $\mu$, associated to the stress tensor, is also completely standard.  Handling its moduli and those of the $b$-ghost amounts to the usual prescription to strip the $c$ ghosts from $n-3$ of the vertex operators and integrate the resulting expression over the moduli space ${\cal M}_{0,n}$ of the $n$-punctured worldsheet.

\subsubsection{The Gaudin Hamiltonian}
\label{sec:Gaudin}

In part, the moduli of the worldsheet gauge field $\tilde{e}$ may be handled similarly. The presence of the $\tilde c$ insertions restrict our BRST transformations to vanish at the punctures, and so they cannot be used to set $\tilde e=0$. If we let $\{\tilde{e}_\alpha\}$ be a basis of $H^{0,1}(\Sigma,T_\Sigma(z_1+\cdots+z_n))$ then we are free to gauge fix $\tilde{e}(z) = \sum_{\alpha} r_\alpha \,\tilde{e}_\alpha(z)$ for some coefficients $r_\alpha$. The path integral over the gauge field $\tilde e(z)$ reduces to an $n-3$ dimensional integral over these coefficients. If  $(\tilde e_\alpha, Q)$ denotes the natural pairing $\int_\Sigma e_\alpha(z) \,Q(z)$ between $e_\alpha\in H^{0,1}(\Sigma,T_\Sigma(z_1+\cdots+z_n))$ and $Q\in H^0(\Sigma,K_\Sigma^2(-z_1-\cdots-z_n))$, then a standard choice for the $\{\tilde{e}_\alpha\}$ gives $(\tilde e_i,Q)= {\rm Res}_{z_i} Q$, with the residue viewed as a section of $K_i(-\sum_{j\neq i} z_j)$. In this case, the path integral over the corresponding $\tilde b\tilde c$ ghost system simply leads to an additional factor of $1/{\rm Vol\,SL(2)}$  (coming from the zero-modes of the $\tilde{c}$s and acting on the worldsheet, not target space). In short, after integrating out all the ghosts, we are left with the integral\footnote{As indicated, since the whole theory has been chiral, the remaining moduli integral should really be interpreted as a contour integral, taken over some $n-3$-dimensional real cycle $\Gamma\subset T^*{\cal M}_{0,n}$.}
\begin{equation}
\label{eqn:holomorphic-gauge-full-correlator}
\mathcal{A}_{+\cdots+} = \int\limits_{\Gamma\subset T^*\!{\cal M}_{0,n}} \!\!\!\!\!\! \frac{\diff z_1\,\diff z_2\,\cdots\,\diff z_n}{({\rm Vol \,SL(2)})^2}  ~\diff^{n-3}r  ~ \sum_{\alpha} {\rm PT}(\alpha)
\frac{1}{z_{12}} \,
\left\langle  {\rm e}^{-(\tilde{e},H)} ~  A^{\prime +}  (z_1) \, A^{\prime +}  (z_2)
    \,  \prod_{i=3}^n {\cal A}^{\prime +}  _i (z_i)
   \right\rangle_0 \,,
\end{equation}
where ${A}^{\prime+}$ is defined by $A^+ = c\tilde c\,A^{\prime+}$ and similarly for ${
\cal A}^{\prime}+$. The subscript on the correlation function here indicates that it is to be evaluated using the matter worldsheet CFT, with  $\mu=\chi=0$ and $\tilde e = \tilde e(r)\equiv \sum r_\alpha\tilde e_\alpha 
	~\in~ H^{0,1}(\Sigma,T_\Sigma(z_1+\cdots+z_n))$ a basis of moduli. 

\medskip

The integral over the  moduli of $\tilde e$ must be handled carefully. As explained in the Introduction, in flat space with plane wave external states, the CFT operator $H$ could be evaluated explicitly in terms of external momenta. Integrating over the moduli parameters $r$ then leads to $\delta$-functions imposing the scattering equation constraints. (See~\cite{Ohmori:2015sha,Mizera:2019gea} for a more rigorous derivation using Picard-Lefschetz theory.)  On AdS$_3\times S^3$, we could formally claim that integrating over the $r$s leads to an insertion of $\prod_{j=4}^n\,\bar\delta({\rm Res}_jH) = \prod_{j=4}^n \partialb (1/{\rm Res}_jH)$, but this is not useful because $H$ is still a CFT operator and we must evaluate the CFT correlation function before we can give meaning to its residues. Parenthetically, a somewhat similar situation would arise in flat space if one tried to compute amplitudes for generic on-shell external states, without using their plane-wave decomposition.

\medskip

To make progress, we need to understand how the current $J$  acts on either the wavefunction  factors $\Phi_2(g) = 1/[\tilde\lambda|g|\lambda\rangle^2$, or the various fields contracted with polarization tensors. To begin, first recall the OPE
\begin{equation}
\label{eqn:J-vertex-OPE}
	J_a(z)\,\Phi(g(z_i)) \sim \frac{1}{z-z_i} (e_a\Phi)(g(z_i)) \,,
\end{equation}
where $\Phi(g(z_i))$ is any scalar function of the field $g(z_i)$ and where $e_a\Phi$ is the right action of the group on $\Phi$. We also recall from~\eqref{eqn:lambda-trans}-\eqref{eqn:lambda-tilde-trans} that the combination $[\tilde\lambda|g(z_i)|\lambda\rangle$ appearing in our bulk-to-boundary propagator is invariant under right translations acting simultaneously on $g(z_i)$ and $|\lambda\rangle$. Thus, for the specific case of bulk-to-boundary propagators of weight $\Delta$, \eqref{eqn:J-vertex-OPE} becomes
\begin{equation}
	J_a(z)\,\Phi_\Delta(g(z_i)) \sim -\frac{1}{z-z_i} ~{\frak t}_a\Phi_\Delta(g(z_i)) \,.
\end{equation}
where the ${\frak t}_a$ acts only on the {\it boundary} point and the change in sign reflects the fact that $|\lambda\rangle$ transforms in the opposite sense to $g(z)$ under right translations. Explicitly, choosing the basis $\{{\frak e},\,{\frak f}\,,{\frak h}\}$ for $\frak{sl}_2$, when acting on a function of homogeneity $-\Delta$, the ${\frak t}_a$ may be described by\footnote{These generators obey the algebra $[{\frak h},{\frak e}] = 2{\frak e}$, $[\frak{h},{\frak f}]=-2{\frak f}$ and $[{\frak e},{\frak f}]={\frak h}$. In the equivalent situation that each vertex operator is multiplied by a factor of $\langle\lambda\,\diff\lambda\rangle\,[\tilde{\lambda}\,\diff\tilde{\lambda}]$ to make it weightless, the $\frak{t}$s should be taken as Lie derivatives on the boundary, and thus also see the boundary forms.}
\begin{equation}
\label{eqn:SL2-generators-on-boundary}
 {\frak h} =  2 x\, \partial_{x} +\Delta ~, \qquad 
 {\frak e} =  -x^2  \partial_{x} - \Delta \, x ~, \qquad 
 {\frak f} = \partial_{x}  ~,
 \end{equation}
for right-translations, with similar expressions for left-translations in terms of $\tilde x$.

Applying this to the vertex operators, we have
\begin{equation}
\label{eqn:J-on-psi-pol}
	J_a(z) \, \langle i |   \psi(z_i) |i\rangle ~\sim ~ - \frac{1}{z-z_i} \,\frak{t}_{ia}  
	\langle i|\psi(z_i)|i\rangle\,,
 \end{equation}
which follows straightforwardly from the $J\psi$ OPE, where $|i\rangle = |\lambda_i\rangle$, $|i]=|\tilde\lambda_i]$, and $\frak{t}_{ia}$ acts on $x_i$ as in~\eqref{eqn:SL2-generators-on-boundary}, here with weight $\Delta=-2$. The fact that $J$ transforms $\psi$ in the adjoint transfers to the  action of $\frak{t}_{i}$ on both factors of $\lambda_i$ in~\eqref{eqn:J-on-psi-pol}. The OPE of $J$ with a polarization structure involving another copy of $J$ needs to be done with more care, because of the double poles in the $JJ$ OPE. Nevertheless, we find that the particular combinations of polarizations and $1/[\tilde\lambda|g|\lambda\rangle$ factors appearing in the vertex operators ${\cal A}^+$ have only simple poles with $J$. In particular
\begin{equation}
 \label{eqn:J-action-on-gauge-vertexops}
  J_a(z) \, \Phi_2^{(i)} \,\epsilon^{(i)}\cdot J (z_i) ~\sim~
   -  \frac{1}{z-z_i} \, \frak{t}_{ia}  \left( \frac{\langle i|J|i\rangle}{[i|g(|i\rangle^2}  \right)\!(z_i)\,,
 \end{equation}
again acting only on the external data. We also note that $J(z)H(w)\sim 0$ so that the ambitwistor constraint $H$ is invariant under $SL(2)$ transformations, even at the quantum level

\medskip

We can use this to understand the factor of ${\rm e}^{-(\tilde e(r),H)}$, acting on the remaining terms in the correlation function. Recalling from~\eqref{eqn:HH-OPE-trivial} that the $HH$ OPE is trivial, we may expand the exponential as a power series without worrying about ordering the factors. Each occurrence of $H= \frac{1}{2} J^2$ may then be handled using the OPEs above. Defining
\begin{equation}
\label{eqn:spectral-sl2}
	\frak{h}(z) = \sum_{i=1}^n \frac{\frak{h}_i}{z-z_i}\,,\qquad
	\frak{e}(z) = \sum_{i=1}^n \frac{\frak{e}_i}{z-z_i}\qquad\text{and}\qquad
	\frak{f}(z) = \sum_{i=1}^n \frac{\frak{f}_i}{z-z_i}
\end{equation}
we have that $H(z)$ acts remaining insertions as the quadratic Casimir\footnote{We abuse notation slightly by continuing to denote this operator by $H(z)$.}
\begin{subequations}
\begin{equation}
\label{eqn:Gaudin-H(z)}
\begin{aligned}
	H(z) ~&=~\frak{e}(z)\frak{f}(z) + \frak{f}(z)\frak{e}(z) + \frac{1}{2}\frak{h}(z)^2\\
		&= \frac{1}{2}\sum_i \frac{\Delta_i(\Delta_i-2)}{(z-z_i)^2} ~+~\sum_{i,j} 
\frac{\frak{e}_i\frak{f}_j + \frak{f}_i\frak{e}_j + \frac{1}{2}\frak{h}_i\frak{h}_j}{(z-z_i)(z-z_j)}\\~&=~ \frac{1}{2}\sum_i \frac{\Delta_i(\Delta_i-2)}{(z-z_i)^2}~ + ~\sum_{i=1}^n \frac{H_i}{z-z_i} 
\end{aligned}
\end{equation}
where 
\begin{equation}
	H_i = \sum_{j\neq i}\frac{\frak{e}_i\frak{f}_j+ \frak{f}_i\frak{e}_j + \frac{1}{2}\frak{h}_i\frak{h}_j}{z_i-z_j}
\end{equation}
\end{subequations}
are the residues ${\rm Res}_i H(z)$. Again, these operators now act only on boundary data. Since $[H(z),H(z')]=0$, the residues themselves obey\footnote{Note that, unlike in the Abelian flat space case, individual {\it summands} of~\eqref{eqn:Gaudin-H(z)} do {\it not} commute.} $[H_i,H_j]$ for all $i,j$, so we may write
\begin{equation}
	{\rm e}^{-\sum_j r_j(\tilde e_j,H)} = {\rm e}^{-\sum_j r_jH_j} 
	= \prod_{j=4}^n {\rm e}^{-r_jH_j}\,,
\end{equation}
and bring this product outside the CFT correlator. 

\medskip

The operator~\eqref{eqn:Gaudin-H(z)} is the Hamiltonian of the Gaudin model, one of the simplest and best-studied quantum integrable systems~\cite{Gaudin:1976sv,GaudinBook,Sklyanin:1987ih}. In the context of spin chains, $H(z)$ acts on the tensor product $\bigotimes_i V_i$ spanned by the spins. For the ambitwistor string, the  worldsheet insertion points play the role of spectral parameters for the spins. Examining the  insertions remaining in the correlator shows that the $V_i$ are principal series representations of ${\frak sl}_2$ spanned by functions of  $|i\rangle$ that are homogeneous of weight $0$.  Crucially, these representations have vanishing quadratic casimir, ensuring the double pole terms in~\eqref{eqn:Gaudin-H(z)} all vanish. This reflects the fact that our vertex insertions were BRST-closed. In addition, since the correlator is invariant under global $SL(2)$ transformations acting diagonally on all the boundary points, the generators~\eqref{eqn:spectral-sl2} are all regular at $z=\infty$, and therefore so too is $H(z)$. In particular, ${\rm Res}_{z_i\to\infty}(H_i)=0$. Invariance under global $SL(2)$ transformations and the massless condition $\Delta_i(\Delta_i-2)=0$ thus ensure $H_i$ are invariant under worldsheet M{\"o}bius transformations. They are exactly analogous to momentum conservation and  masslessness in flat space.

\subsubsection{Completing the evaluation of the worldsheet correlator}
\label{sec:completing-the-all-plus}

Using the OPEs above, we can also eliminate all the $J$ insertions from the vertex operators ${\cal A}^{\prime+}$, again trading them for operators acting on the boundary data. As with the Gaudin Hamiltonian, the virtue of this is that such boundary operators may be brought outside the CFT correlation function. Let us define 
\begin{equation}
\epsilon_i \cdot \mathcal{T}_i  \equiv \sum_{j \neq i} \frac{ \epsilon _i \cdot \frak{t}_j}{z_{ij}}  = \sum_{j \neq i} \frac{ \langle i | \frak{t}_j | i \rangle  }{z_i-z_j} = \sum_{j \neq i} \frac{ x_{ij} }{z_{ij}}  \left( \Delta_j - x_{ij} \, \frac{\partial}{\partial x_j}  \right)\,
\end{equation}
where $z_{ij} = z_i-z_j$ and similarly for $x_{ij}$. Then we find that the result of performing all OPEs involving all $J$ insertions in the correlator
 \begin{equation}
\label{eqn:gauge-corr-2}
\left\langle {\rm e}^{-(\tilde e,H)} {A}^{\prime+}_1{A}_2^{\prime+}\,
\prod_{k=3}^n {\cal A}^{\prime+}_k \right \rangle_0  =    
\prod_{j=4}^n {\rm e}^{-r_jH_j} \,\left[ 
  \left( \prod_{i=3}^n  \epsilon_i  \! \cdot \!  \mathcal{T}_i \right) \frac{\epsilon_1  \! \cdot \!  \epsilon_2}{z_{12}}
  ~ \left\langle \prod_{k=1}^n \Phi_2(g(z_i))\right\rangle_0\right]\,,
\end{equation}
with the ${\cal T}$s acting on the result of the remaining correlation function.  The factor of $\epsilon_1\cdot\epsilon_2 / z_{12} = \langle 12\rangle^2/z_{12}$ arose from performing the $\psi$ contraction once all the $J$s had been removed. The differential operator sitting outside the correlation function should be thought of as analogous to the CHY Pfaffian that enters the flat space amplitudes (for this polarization configuration).

The remaining  worldsheet CFT correlator is simple to perform. Since the only remaining field $g(z)$ has trivial OPE $gg\sim0$, the correlator must be regular everywhere. As it is a scalar on $\Sigma_i$, it must in fact be independent of all the $z_i$s and so reduces to an integral over the zero-mode of $g$. This yields\footnote{In the full model, there will also be a factor of ${\rm Vol}({\rm S}^3\times M$.} the well-known D-function~\cite{DHoker:1999kzh} 
\begin{equation}
\label{eqn:D-function-general}
	{\rm D}_2(x_i) ~=~ {\rm D}_{2,\ldots,2}(x_1,x_2,\ldots,x_n)~=~ \int_{{\rm AdS}_3} \prod_{i=1}^n \frac{1}{[i|g|i\rangle^2}~\diff^3g\,,
\end{equation}
here of weight $2$ in each of the $n$ boundary points. The D-function represents a contact interaction in AdS and may be viewed as the AdS analogue of the momentum-conserving $\delta$-function in flat space. 

\medskip

To summarize, with $n$ holomorphic gluons, the worldsheet path integral may be evaluated as
\begin{equation}
\label{eqn:gauge-holomorphic-matter-correlator}
\int\limits_{\Gamma\subset T^*\!{\cal M}_{0,n}} \!\!\!\!\!\! \frac{\diff z_1\,\cdots\,\diff z_n~\diff^{n-3}r}{({\rm Vol \,SL(2)})^2}  ~ \frac{1}{z_{12}} \,\left(\sum_{\alpha \in S_n/D_n} \!\!{\rm PT}(\alpha)\right)\,
  \left( \prod_{i=3}^n  \epsilon_i  \! \cdot \!  \mathcal{T}_i \right) \,\frac{\epsilon_1  \! \cdot \!  \epsilon_2}{z_{12}}\,\prod_{j=4}^n{\rm e}^{ - r_jH_j }  \,
  {\rm D}_{2}(x_i)\,.
\end{equation}
where the sum is over dihedrally  inequivalent permutations of $\{1,2,\ldots,n\}$. While our derivation placed the Gaudin Hamiltonian as acting from the far right, we have checked that, when acting ultimately on a D$_2$-function, in fact the $H_j$ commute with the $\epsilon_i\cdot{\cal T}_i$s, with the only difference being the operators are always involve the appropriate weight factors $\Delta_i$ for whatever sits to their right. The Gaudin Hamiltonian  now acts directly on a D-function ${\rm D}_2(x_i) = {\rm D}_{2,\ldots,2}(x_1,\ldots,x_n)$ of weight 2 in each  boundary point. This weight again ensures that the Gaudin Hamiltonian has no double poles.

\medskip

Let us make some remarks on this expression. The correlator \eqref{eqn:hol-gauge-corr-undescended} is manifestly symmetric under permutations of the labels $4, \cdots , n$, and so the expression~\eqref{eqn:gauge-holomorphic-matter-correlator} must also possess this symmetry. To see that it does, observe that
\begin{equation}
\left [ \,\epsilon_i \cdot \mathcal{T}_i \,,\,  \epsilon_j \cdot \mathcal{T}_j \,\right] 
= \frac{1}{z_{12}^2}  \,\epsilon_i ^a \, \epsilon_j ^b  \, {\rm f}_{abc}  \left(\frak{t}_i^c + \frak{t}_j^c \right) ~.
\end{equation}
 The operator on the right vanishes inside~\eqref{eqn:gauge-holomorphic-matter-correlator}, by virtue of the $\frak{sl}_2$ identity 
\begin{equation}
\label{eqn:antisymmetric-derivative-identity}
	\epsilon_{[a} e _{b]} \, \Phi_\Delta = - \frac{\Delta}{4}  \,  f_{ab}^c \,  \epsilon _{c}  \, \Phi_\Delta\,.
\end{equation}
Secondly, just as global boundary $SL(2)$ invariance ensured the operators~\eqref{eqn:spectral-sl2} had no pole at $z=\infty$, so too the combinations $\epsilon_i \cdot \mathcal{T}_i $ are well-defined meromorphic differential operators. For example, we have
 \begin{equation}
 \mathrm{Res}_{z_i \to \infty} (\epsilon_i \cdot \mathcal{T}_i)  ~= ~
 \sum_{j \neq i}  \epsilon_i \cdot \frak{t}_j \,=\,- \epsilon_i \cdot \frak{t}_i \,.
 \end{equation}
With the $\frak{t}_i$ represented as in~\eqref{eqn:SL2-generators-on-boundary}, this expression vanishes identically. Thus we have verified that the integrand in \eqref{eqn:gauge-holomorphic-matter-correlator} is indeed a meromorphic quadratic differential in the $z_i$, with the appropriate permutation symmetry. In the case $n=3$ the factor ${\rm e}^{-r_jH_j}=1$ and the D-function is simply
 \begin{equation}
 {\rm D}_{2,2,2}(x_1,x_2,x_3) \propto  \frac{1}{\langle12\rangle\langle23\rangle\langle31\rangle}\, \frac{1}{[12][23][31]}
 \end{equation}
As a further small consistency check on our differential operators $\epsilon_i\cdot{\cal T}_i$, acting on  this expression with $\epsilon_3\cdot \mathcal{T}_3$ cancels all the dependence on the $|\lambda\rangle$s, leaving us with 
 \begin{equation}
 \mathcal{A}_{+++}(x_1,x_2,x_3) \propto \frac{1}{[12][23][31]}\,.
 \end{equation}
Of course, this is the correct answer for the correlator $\langle \bar{J}(x_1) \bar{J}(x_2) \bar{J}(x_3) \rangle$ of three antiholomorphic currents in the dual boundary CFT.

\subsection{General gluon amplitudes}
\label{sec:general-gluons}

Next we turn our attention to the general correlator of $n$ gluons, with arbitrary numbers of each polarization. This correlator can again be computed by repeated application of the $J$  OPEs, and we shall see that the result is very similar to~\eqref{eqn:gauge-holomorphic-matter-correlator}.

\medskip

Many of the steps are similar to those above, so we will be brief, highlighting only the differences. We arbitrarily choose the picture $-1$ gluons to be of the holomorphic type. Integrating out the ghosts as before, the correlator we require is
\begin{equation}
\left\langle 
 {\rm e}^{-(\tilde{e},H)} ~  A^{\prime +}_1 (z_1) \, A^{\prime +}_2 (z_2)
    \,  \prod_{i\in +} {\cal A}^{\prime +} _i (z_i)\,\prod_{j\in -}{\cal A}^{\prime -}(z_j)
   \right\rangle _0 \,
\end{equation}
The difference to the previous case is that the polarization structure of the ${\cal A}^-$s depends on the field $g$. Remarkably, we nonetheless find the current $J$ has the OPE
\begin{equation}
	J_a(z)\,:\!\Phi_2\,\bar\epsilon^b(g)J_b\!:(w) ~\sim~ - \frac{\frak{t}_a}{z-w}  :\!\Phi_2\,\bar\epsilon^b(g)J_b\!:(w)
\end{equation}
with the ${\cal A}^-$ vertex operator, where again the $\frak{t}$ acts only on the $|\lambda\rangle$ dependence of this operator. We caution the reader that to obtain this result, it is crucial to order the vertex operator in the way we have given.

This result is important. Firstly, it means that the factors of $J$ we meet in expanding ${\rm e}^{-(\tilde e(r),H)}$ can be handled just as in the all holomorphic polarization case. In particular, we can pull the Gaudin Hamiltonian outside the correlation function exactly as before. Similarly, all factors of $J$ in the remaining holomorphic vertex operators may similarly be replaced by the classical differential operators $\epsilon_i\cdot{\cal T}_i$, again coming outside the correlator. 

Finally, we consider the $J$s that are attached to the field-dependent antiholomorphic polarization vectors. At this stage it is important that there are no remaining insertions of $\epsilon\cdot J$. We find the result
\begin{equation}
\label{eqn:antihol-J-acts}
	\bar\epsilon(g)\cdot J(z) \,:\!\Phi_2\,\bar\epsilon^b(g)J_b\!:(w)~\sim~ 
	- \frac{\bar\epsilon\cdot\bar{\frak{t}}}{z-w}  :\!\Phi_2\,\bar\epsilon^b(g)J_b\!:(w)
\end{equation}
where $\bar{\frak{t}}$ are the generators~\eqref{eqn:SL2-generators-on-boundary}, but acting on  $|\tilde\lambda]$ (or $\tilde x$). Note that since $|\lambda\rangle$ and $|\tilde\lambda]$ are treated as independent complex variables, $[\,\frak{t}_a,\bar{\frak{t}}_b\,]=0$ for all pairs of generators. 

\medskip

Performing the remaining $\psi$ contractions and reducing the $g$ path integral to a D-function as before, the heterotic ambitwistor string yields the formula
\begin{equation}
\label{eqn:gauge-general}
\begin{aligned}
	{\cal A}_n^{1\tr}~ &= ~\int \frac{\diff z_1\,\cdots\,\diff z_n~\diff^{n-3}r}{({\rm Vol \,SL(2)})^2}  ~ \frac{1}{z_{12}} \,\left(\sum_{\alpha} {\rm PT}(\alpha)\right)\\
	&\qquad\qquad\times~\left[
	\prod_{i\in +\backslash \{1,2\}}  \epsilon_i  \! \cdot \!  \mathcal{T}_i  ~
	 ~\frac{\epsilon_1\cdot\epsilon_2}{z_{12}}~
 \prod_{j\in -} \epsilon_j\!\cdot \! \bar{{\cal  T}}_j~
 \prod_{j=4}^n{\rm e}^{ - r_jH_j } ~\right]\, {\rm D}_{2}(x_i)
\end{aligned}
\end{equation}
as the complete tree-level scattering of $n$ gluons in Yang-Mills-Chern-Simons theory on AdS$_3$. The Gaudin Hamiltonians may again be brought through the other operators to act directly on the D-function. Again, the ordering of the $\epsilon_i\cdot{\cal T}_i$ and $\bar{\epsilon}_j\cdot\bar{\cal T}_j$ operators is immaterial, and $\bar{\epsilon}_j\cdot\bar{{\cal T}}_j$ is defined exactly as $\epsilon_i\cdot{\cal T}_i$, except using the 
$\bar{\frak t}_i$ generators acting on the antiholomorphic boundary coordinates. (All worldsheet dependence is still holomorphic.)

\subsection{Correlator of $n$ holomorphic gravitons}
\label{sec:n-holomorphic-gravitons}

We now consider gravitational amplitudes in the type II model. In this paper, we will content ourselves with the simplest case of all holomorphic polarizations:
\begin{equation}
\label{eqn:hol-corr-undescended}
\mathcal{M}_{++\cdots+}(x_1,\ldots,x_n) = \left\langle U_1^+(z_1)\,U_2^+(z_2)  \prod_{i=3}^n \,V_i^+(z_i) \right\rangle \,.
\end{equation}
In this expression, $V^+$ is the picture zero graviton operator, which may be computed from the OPE of the PCOs $\delta(\beta)\,G$ and $\delta(\tilde\beta)\tilde G$ with $U^+$. We find
\begin{equation}
\label{eqn:graviton-picture-zero}
\begin{aligned}
	V^+  & = c\,\tilde c~ 
	\left(\epsilon \!\cdot\!   j -  \psi^a\psi^b\epsilon_a\, e_b \right) 
	\left(\epsilon \!\cdot\!   j - (\psi^c\psi^d +\psit^c\psit^d)\epsilon_c\, e_d\right) 
	  \, 	\frac{1}{[\tilde\lambda|\,g\,|\lambda\rangle^4}\\
	  &= c\,\tilde c~
	\frac{\langle\lambda|\,J\,|\lambda\rangle \,\langle\lambda|\,K\,|\lambda\rangle}{[\tilde\lambda|\,g\,|\lambda\rangle^4}\,,
\end{aligned}
\end{equation}
where in the first line, $j$ is related to $J$ as in~\eqref{eqn:new-J}, $\epsilon$ is the polarization vector and $e_a$ is the right derivative operator on the group, acting on the wavefunction factor $1/[\tilde\lambda|g|\lambda\rangle^4$. The first line takes a form familiar from flat space. To reach the second line, we used the $\frak{sl}_2$ identity~\eqref{eqn:antisymmetric-derivative-identity} and introduced the shorthand $K_a \equiv J_a + \frac{1}{2} \mathrm{f}_{\,abc}\,\psi^b\psi^c$. This packaging will be convenient in what follows. While $J$ absorbed just one set of fermions, $K$ absorbs both fermion and  in consequence obeys a level $-1$ current algebra, with OPEs
\begin{equation}
\label{eqn:K-OPEs}
\begin{gathered}
	K_a(z)\,K_b(w) ~\sim ~J_a(z)\,K_b(w) ~\sim~ \frac{-\kappa_{ab}}{(z-w)^2} + \f abc \frac{K_c(w)}{z-w}\,,\\
	K_a(z) g(w) ~\sim~ \frac{g t_a}{z-w} \,,\qquad K_a(z)\psi^b(w)~\sim ~K_a(z)\psit^b(w)~\sim~0
\end{gathered}	
\end{equation}
among the other fields. Again, it is remarkable that the picture zero vertex operator of a string theory on a curved background can take such a simple form. With $n-2$ vertex operators raised to picture zero, intergrating over the zero mode of both $\gamma$ and $\tilde\gamma$ leads to a factor of $1/z_{12}^2$. 

To do this, we first note that, with the currents packaged as $J$ and $K$,  $\psit$ has trivial OPE with all the other fields. It may therefore be integrated out immediately, contracting one set of  polarization tensors in $U_1^+\, U_2^+$ and giving a factor of $\epsilon_1\cdot\epsilon_2/z_{12}$. We now turn to the $J$s in the polarization tensors. In addition to~\eqref{eqn:J-on-psi-pol}, we need the OPEs
\label{eqn:J-action-on-vertexops}
 \begin{equation}
 \begin{aligned}
  J_a(z) \, \left(\frac{\langle i|K|i\rangle}{[ i|g|i\rangle^4}  \right)\!(z_i) ~&\sim~
   -  \frac{1}{z-z_i} \, \frak{t}_a  \left( \frac{\langle i|K|i\rangle}{[i|g(|i\rangle^4}  \right)\!(z_i)\,,\\
   J_a(z) \, \left(\frac{\langle i|J|i\rangle \,\langle i|K|i\rangle}{[i|g|i\rangle^4}  \right) \! (z_i)~&\sim~ -  \frac{1}{z-z_i} \, \frak{t}_a  \left( \frac{\langle i|J|i\rangle \,\langle i|K|i\rangle}{[i|g|i\rangle^4}  \right) \!  (z_i)\,.
 \end{aligned}
 \end{equation}
Again, we find that, despite the $JJ$ and $JK$ OPEs each having double poles, the particular combinations of polarizations and $1/[\tilde\lambda|g|\lambda\rangle$ factors appearing in the vertex operators $V^+$ have only simple poles with $J$. As before, these facts allow us to trade all factors of $J$, whether in the polarization tensors or $H$, for boundary differential operators. 

In the case of gravity, it proves convenient to bring the polarization structure   outside the correlator before Gaudin Hamiltonian. Doing so leaves us with the remaining matter correlator
 \begin{equation}
\label{eqn:grav-holomorphic-matter-correlator-intermediate}
  \frac{\epsilon_1  \! \cdot \!  \epsilon_2}{z_{12}}
  \left( \prod_{i=3}^n  \epsilon_i  \! \cdot \!  \mathcal{T}_i \right) 
  ~ \left\langle {\rm e}^{ - (\tilde{e},  H )}  \prod_{j=1}^2
\,\frac{\langle j|\psi(z_j)|j\rangle}{[j|g(z_j)|j\rangle^4}\,\prod_{k=3}^n  \frac{\langle k| K|k\rangle}{[k|g|k\rangle^4}(z_k)\right\rangle_0
\end{equation}
Since $K\psi\sim0$, performing the $\psi$ path integral simply gives and additional factor of $\epsilon_1\cdot\epsilon_2 / z_{12}$. Finally, we can perform all the $K$ OPEs. 
Since there are no fermions or $J$ insertions remaining,  $K$ acts as
\begin{subequations}
\begin{equation}
K_a(z)\,\Phi_\Delta(g(z_i)) ~\sim~ -\frac{\frak{t}_{ai}\Phi_\Delta (g(z_i))}{z-z_i}
\end{equation}
on scalar functions of the fields, and
\begin{equation}
 K_a(z) \, \left(\frac{\langle i |K|i\rangle}{[i|g|i\rangle^4}  \right) \! (z_i) ~ \sim~ 
 -  \frac{1}{z-z_i} \, \frak{t}_{ai}  \left( \frac{\langle i|K|i\rangle}{[i|g|i\rangle^4}\right)  \! (z_i)\,,
\end{equation}
\end{subequations}
with this specific combination having no double pole. As always, these ${\frak t}$s act on the boundary data with the appropriate weights. Performing all the $K$ OPEs thus gives us a second set of polarization operators. The final $g$ path integral now evaluates to a D$_4$-function.

\medskip

Combining all the ingredients, the ambitwistor string yields the tree-level scattering amplitude for $n$ holomorphic gravitons on AdS$_3$ as
\begin{equation}
\label{eqn:grav-holomorphic-matter-correlator}
 \int\limits_{\Gamma\subset T^*\!{\cal M}_{0,n}} \!\!\!\!\!\! \frac{\diff z_1\,\cdots\,\diff z_n~\diff^{n-3}r}{({\rm Vol \,SL(2)})^2}  ~ \frac{1}{z_{12}^2} \,\frac{\epsilon_1  \! \cdot \!  \epsilon_2}{z_{12}}
  \left( \prod_{i=3}^n  \epsilon_i  \! \cdot \!  \mathcal{T}_i \right) 
  ~ \prod_{j=4}^n{\rm e}^{ - r_jH_j }  
  ~ \frac{\epsilon_1  \! \cdot \!  \epsilon_2}{z_{12}} 
  \left( \prod_{k=3}^n  \epsilon_k  \! \cdot \!  \mathcal{T}_k \right)   {\rm D}_4(x_i)\,.
\end{equation}
While this may look intimidating, we emphasize that all quantum path integrals have now been performed. We are left with a classical, finite-dimensional integral. In particular, the operators $H_j$ are the classical differential operators of~\eqref{eqn:Gaudin-H(z)}.

\subsection{Bi-adjoint scalar amplitudes}
\label{sec:bi-adjoint-scalars}

We also briefly consider the bi-adjoint scalar theory, with action
\begin{equation}
\label{eqn:bi-adjoint-scalar-action}
	S[\Phi] = \int_{{\rm AdS}_3} ~\frac{1}{2}\,D\Phi_{AA'} \wedge *\,D\Phi^{AA'} \,+\, \frac{1}{3}\, {\rm f}_{ABC}\, {\rm f}'_{A'B'C'}\, *\Phi^{AA'}\,\Phi^{BB'}\,\Phi^{CC'}
\end{equation}
on AdS$_3$.  This theory is expected to come from an ambitwistor string where both $\psi$ and $\psit$ are replaced by auxiliary worldsheet current algebras, with structure constants ${\rm f}^A_{BC}$ and ${\rm f}^{A'}_{B'C'}$. This scalar theory has no polarization structure and is certainly non-chiral, so to obtain the correct amplitudes we should use the Hamiltonian $H(z) + \bar H(z)$, where $H(z)$ is as in~\eqref{eqn:Gaudin-H(z)} and $\bar{H}(z)$ is similar, but with the generators~\eqref{eqn:SL2-generators-on-boundary} replaced by generators $\{\bar{\frak{h}},\bar{\frak{e}},\bar{\frak{f}}\}$ that act instead on the $\tilde x$s. In particular, $\{\frak{t},\bar{\frak{t}}\}=0$ for any pair of holomorphic and antiholomorphic boundary generators. Note that  $H$ and $\bar{H}$ both depend only on the holomorphic worldsheet coordinate.

The double-leading trace contribution to $n$-point amplitudes in the theory~\eqref{eqn:bi-adjoint-scalar-action} will then be given simply by
\begin{equation}
\label{eqn:scalar-amplitude}
{\cal A}^{\rm scal}
= \int\limits_{\Gamma\subset T^*\!{\cal M}_{0,n}} \!\!\!\!\!\! \frac{\diff z_1\,\cdots\,\diff z_n~\diff^{n-3}r}{({\rm Vol \,SL(2)})^2}  ~ \sum_{\alpha,\beta\in S_n/D_n}{\rm PT}(\alpha)\,{\rm PT}(\beta)~\prod_{j=4}^n{\rm e}^{ - r_j(H_j +{\bar H}_j)}  \,
  {\rm D}_{2}(x_i,\tilde x_i)\,.
\end{equation}
with two Parke-Taylor factors and the same D$_2$-function as in the gauge theory case.

We can also combine the action of $H+\bar H$ into a single term as follows. Let $X = |\tilde\lambda]\langle\lambda|$ be the embedding space representative of the boundary coordinate, obeying $X^2=0$ and $X\sim rX$. Since the D$_2$-function depends on the $(x_i,\tilde x_i)$ only through $X_i$, we can use the representation ${\cal D}_i = X_i^\mu (\partial/\partial X_i)^\nu - X_i^\nu (\partial/\partial X_i)^\mu$ of $SL(2)$ as Lorentz transformations in the embedding space to write
\[
	\frak{t}_i\cdot\frak{t_j} + \bar{\frak{t}}_i\cdot\bar{\frak{t}}_j = 	{\cal D}_i\cdot{\cal D}_j 
\]
Thus, each pair of  Gaudin Hamiltonians becomes
\begin{equation}
	H_i + \bar{H}_i = \sum_{j\neq i} \frac{{\cal D}_i\cdot{\cal D}_j}{z_i-z_j}
\end{equation}
We note that this description extends immediately to higher dimensional AdS.

\section{The scattering equations on AdS}
\label{sec:scattering-equations}

In the previous section, the ambitwistor string was shown to transform the problem of computing AdS scattering amplitudes into the problem of evolving the worldsheet correlator through Euclidean times $r_j$ using the operators ${\rm e}^{-r_jH_j}$. This amounts to finding the mutual eigenfunctions and eigenvalues of the $H_j$, and expanding the result of our worldsheet correlator, either a D-function or some derivatives of a D-function, in this basis.

\medskip

Knowing the set of eigenvalues of the $H_i$ is the same thing as knowing the eigenvalue $\tau(z)$ of $H(z)$ in~\eqref{eqn:Gaudin-H(z)}. This operator transforms as a meromorphic quadratic differential in $z$, having at most simple poles at the $z_i$. Furthermore,  the residue $H_i={\rm Res}_i\, H(z)$ itself has simple poles at all the $z_j$ with $j\neq i$, so transforms as a holomorphic section of $K_{\Sigma_i}\!\left(-\sum_{j\neq i}z_j\right)$. Since $\{\frak{h}_i,\frak{e}_i,\frak{f}_i\}$ do not act on the worldsheet coordinates, under a worldsheet $SL(2)$ transformation, $H(z)\phi(z;x)$ will transform simply as the product of $H$ and $\phi$. Hence the eigenvalue $\tau(z)$ must transform like $H(z)$ itself.  The most general quadratic differential with these properties can be written as
\begin{subequations}
\begin{equation}
\label{eqn:Gaudin-eigenvalue}
	\tau(z) = \sum_{i,j} \,\frac{\tau_{ij}}{(z-z_i)(z-z_j)}\,,
\end{equation}
where the complex parameters $\tau_{ij}$ obey 
\begin{equation}
\label{eqn:parameter-conditions}
	\tau_{ii}= 0\,,\qquad \tau_{ij}=\tau_{ji}\,,\qquad
	\sum_{j\neq i} \tau_{ij} = 0\quad\text{for $i=1,\ldots,n$}\,.
\end{equation}
\end{subequations}
The conditions~\eqref{eqn:parameter-conditions} ensure that $\tau(z)$ has at most simple poles, and that it is non-singular at $z=\infty$  for arbitrary values of the $z_i\in\mathbb{C}$. This ensures we are only considering states that are invariant under global conformal transformations of the boundary.  We argued above that this was true for our CFT correlator. Consequently, the eigenvalues of $H(z)$ must take the form~\eqref{eqn:Gaudin-eigenvalue} and so are parametrized by the $n(n-3)/2$ complex numbers $\tau_{ij}$.  Similarly, ${\rm Res}_i\,\tau(z)$ are the eigenvalues of  $H_i = {\rm Res}_i H(z)$.

Now, let ${\bf z}$ denote the set of $n$ worldsheet points, ${\bf x}$ the boundary points. Also let $\phi_\tau({\bf z};{\bf x})$ be a simultaneous eigenfunction of all the $H_j$s, with corresponding eigenvalues ${\rm Res}_j\tau(z)$. Once the worldsheet correlator is expanded in the $\phi_\tau({\bf z};{\bf x})$ basis, integrating over the Gaudin times $r_j$ with an appropriate contour~\cite{Ohmori:2015sha} will lead to a product of $n-3$ $\delta$-functions, imposing 
\begin{equation}
\label{eqn:scattering-equations}
	\sum_{j\neq i} \frac{\tau_{ij}}{z_i-z_j} = 0
\end{equation}
for any $n-3$ choices of $i\in\{1,\ldots,n\}$. These are our AdS scattering equations. The $n-3$ conditions~\eqref{eqn:scattering-equations} are equivalent to the statement that 
$\tau(z)=0$ identically for all $z\in\Sigma$, and we will often abbreviate them this way.

Exactly as in flat space, the integral over the worldsheet moduli space ${\cal M}_{0,n}$ forces the $z_i$ to sit on a solution to~\eqref{eqn:scattering-equations}.  For example, suppose the result of acting with one set of polarization derivatives on the D-function ${\rm D}_4({\bf x})$ can be written in the $\phi_\tau({\bf z};{\bf  x})$ basis as
\begin{subequations}
\begin{equation}
\label{eqn:worldsheet-correlator-expansion}
	\frac{\epsilon_1  \! \cdot \!  \epsilon_2}{z_{12}} 
  \left( \prod_{k=3}^n  \epsilon_k  \! \cdot \!  \mathcal{T}_k \right) {\rm D}_4({\bf x}) 
  = \int C_\tau({\bf z})\,\phi_\tau({\bf z};{\bf x}) ~ D\tau\,
\end{equation}
and the ${\rm D}_2$ function itself can be expanded as
\begin{equation}
	\frac{\epsilon_1  \! \cdot \!  \epsilon_2}{z_{12}} 
  \left( \prod_{k=3}^n  \epsilon_k  \! \cdot \!  \mathcal{T}_k \right) {\rm D}_4({\bf x}) 
  = \int C^\prime_\tau({\bf z})\,\phi_\tau({\bf z};{\bf x}) ~ D\tau\,
\end{equation} 
\end{subequations}
using a suitable integration measure $D\tau$ over the $n(n-3)/2$ parameters $\tau_{ij}$. Then the ambitwistor string yields the $n$-point $++\cdots+$ graviton amplitude~\eqref{eqn:grav-holomorphic-matter-correlator} and $n$-point $+\cdots+$ gauge theory amplitude~\eqref{eqn:gauge-holomorphic-matter-correlator} in the forms
\begin{subequations}
\begin{gather}
{\cal M}_{+\cdots+}(x_1,\ldots,x_n) ~=~\int \left[~\sum_{{\bf z} \in S}\,
\frac{C_\tau({\bf z})}{{\rm Jac}({\bf z})} \,
\frac{\epsilon_1  \! \cdot \!  \epsilon_2}{z_{12}}~\prod_{i=3}^n \epsilon_i\!\cdot\!  
\mathcal{T}_i   ~ \phi_\tau({\bf z};{\bf x})\right]_{{\bf z}_*} ~D\tau\,,
\label{eqn:gravity-amplitude}\\
{\cal  A}_{+\cdots +} (x_1,\ldots,x_n)  ~=~ \int\left[~\sum_{{\bf z} \in S}\,\sum_{\alpha\in S_n/D_n}\,{\rm  PT}(\alpha)
~\frac{C'_\tau({\bf z})}{{\rm Jac}({\bf z})} \,\phi_\tau({\bf z},{\bf x})\right]_{{\bf z}_*} \,.
\label{eqn:gauge-amplitude}
\end{gather}
\end{subequations}
Similarly, the $n$-point scalar amplitudes can be written as
\begin{equation}
\begin{aligned}
{\cal A}^{\rm scal}(x_1,\ldots,x_n) 
~&=~\int\left[~\sum_{{\bf z} \in S}\,\sum_{\alpha,\beta\in S_n/D_n}
\,{\rm  PT}(\alpha)\,{\rm PT}(\beta)~\frac{C^{\prime\prime}_\tau({\bf z})}{{\rm Jac}({\bf z})}\,\Phi_\tau({\bf z},{\bf x})\right]_{{\bf z}_*} 
\end{aligned}
\end{equation}
in terms of the expansion of ${\rm D}_2(x_i)$ in terms of the eigenstates of the left-right symmetric Gaudin Hamiltonian. In these expressions, Jac$({\bf z}_*)$ is the usual Jacobian\footnote{We understand Jac$({\bf z}_*)$ to include both the Jacobian $\det(\partial_j\tau_i)$ that arises from solving the $\bar\delta(\tau_i({\bf z}))$ constraints, and the Jacobians that arise from fixing the two worldsheet SL$(2;\mathbb{C})$ factors. For the gravity and gauge theory amplitudes, they also include a factor of $1/z_{12}^2$ (gravity) and $1/z_{12}$ (gauge theory) originating from the zero modes of the $\gamma$ and $\tilde\gamma$ ghosts.}  and $S$ is the set of solutions to the $n-3$ scattering equations $\tau_i({\bf z})=0$. We emphasize that these scattering equations take the same form in all three cases. Only the expansion coefficients are different.

\medskip

It is remarkable that the solutions to the AdS scattering equations~\eqref{eqn:scattering-equations} take exactly the same form as in flat space, with the replacement $s_{ij}\to\tau_{ij}$. In particular, the AdS scattering equations also have $(n-3)!$ solutions~\cite{Dolan:2013isa,Dolan:2014ega}. The integral over the $\tau_{ij}$ in~\eqref{eqn:scalar-amplitude2}  is somewhat akin to the integration over momenta one would perform if using the flat-space CHY expressions to evaluate scattering of generic on-shell states, with the important difference that here the particular wavepacket $C_\tau({\bf z}_*)$ generically depends on the solution to the scattering equations. The expansion~\eqref{eqn:worldsheet-correlator-expansion} is also reminiscent of the expansion of a correlation function in terms of conformal partial waves~\cite{Dolan:2003hv,Dolan:2011dv}. The conditions~\eqref{eqn:parameter-conditions} obeyed by the $\tau_{ij}$ are also very suggestive of Mellin parameters. We will find evidence that  the $\tau_{ij}$ are indeed determined by the Mellin parameters. In this sense, the eigenfunctions $\phi_\tau({\bf z}_*;{\bf x})$ are the `basic' objects for AdS scattering processes.

\medskip

We now turn to a preliminary investigation of the eigenfunctions themselves. 

\subsection{Eigenfunctions of the Gaudin Hamiltonians}
\label{sec:eigenfunctions}

The simplest way to find the eigenstates of the Gaudin Hamiltonian uses the algebraic Bethe ansatz~\cite{Gaudin:1976sv,GaudinBook,Sklyanin:1987ih,Sklyanin:1995bm,Sklyanin:1997zz}, which relies on the existence of a suitable vacuum state.  Whether for gravity, gauge theory and the bi-adjoint scalar, we have seen that the Gaudin Hamiltonian always acts on an object of weight 2 in each state, whether this is a simple D$_2$-function or some derivatives of a D$_4$-function. Consequently, we need to understand the eigenstates of a homogeneous spin chain with $\Delta_i=2$. The ABA can be applied to this case, with the vacuum represented by $ 1/(x_1^2x_2^2\cdots x_n^2)$. However, in this representation, the set of Bethe states is known to be incomplete~\cite{Faddeev:1994zg,Derkachov:2001yn,Derkachov:2002wz} and do not form a basis in which we can expand our worldsheet correlators\footnote{The eigenvalue $\tau(z)$ of states built from the Bethe ansatz take the form
\begin{equation}
\label{eqn:Bethe-eigenvalue}
	\tau(z) = \frac{1}{2}\sum_i \frac{\Delta_i(\Delta_i-2)}{(z-z_i)^2}  +{\sum_{i,j}}'\frac{2\Delta_i\Delta_j}{(z-z_i)(z-z_j)}
\end{equation}
provided the Bethe equations hold. If $\Delta_i=2$ then this $\tau(z)$ has no double poles for $z\in\mathbb{C}$, but has a both double and single pole at $z=\infty$.  This is not a problem for the integrable spin chain where the spectral parameter $z\in\mathbb{C}$. However, the singularity at  $z=\infty$ means that none of these states could have arisen from any well-behaved string theory with $z\in\mathbb{CP}^1$.}. We must therefore proceed differently.

\medskip

In this paper, we will only  consider the simplest case of $n=4$ where there is only one scattering equation. The Gaudin eigenfunction problem $(H(z)-\tau(z))\phi_\tau({\bf z};{\bf x}) =0$ reduces to a single differential equation, with the boundary cross-ratio $X= x_{12}x_{34} / x_{23}x_{41}$ being the variable and the worldsheet cross-ratio $Z=z_{12}z_{34}/z_{23}z_{41}$ a free parameter. We obtain
\begin{equation}
\label{eqn:Heun}
\left[\frac{d^2}{dX^2} + \left(\frac{1}{X}+\frac{1}{X-1} +\frac{3}{X-Z}\right)\frac{d}{dX} + \frac{4X-2-\tau(Z)}{X(X-1)(X-Z)}\right]\phi_{\tau}({\bf z};{\bf x}) = 0
\end{equation}
where $\tau(Z)= (Z-1)\tau_{12} + Z\tau_{13}$. Equation~\eqref{eqn:Heun} is a  particular case of Heun's equation -- the general ode having at most four regular singular points $\{0,1,Z,\infty\}\in\mathbb{CP}^1$ and a natural generalization of the hypergeometric equation (see {\it e.g.}~\cite{DigitalLibrary,RonveauxBook}). In general the solutions are rather complicated functions, known as `local Heun functions'. The coefficients of the singularities that arise in our case imply that around the singular points $0,1$, one of the two solutions to~\eqref{eqn:Heun} may be expressed as a (generically infinite) power series beginning with a constant, whilst the other has a logarithmic singularity. For example, in a neighbourhood of $X=0$, the two solutions admit expansions
\[
	\phi^+_{\tau}({\bf z};{\bf x}) ~=~ \sum_{k=0}^\infty a_k(Z,\tau)\, X^k
	\quad\text{and}\quad
	\phi^-_{\tau}({\bf z};{\bf x}) ~=~ \phi^+_{\tau}({\bf z};{\bf x}) \,\ln(X) + \sum_{k=1}^\infty b_k(Z,\tau)\,X^k\,,
\]
where for example the coefficients $a_k(Z,\tau)$ obey the recursion relation
\[
	Z(k+1)^2\,a_{k+1} - (Z\,k(k+1) +k(k+3) +\tau(Z))\,a_k + (k+1)^2\,a_{k-1} ~=~0\,,
\]
with $a_1 = -(\tau(Z)/Z)\, a_0$. It is also possible to reassemble these expansions in terms terms of $_2F_1(k+1,-k;1;X)$ yielding their expansions around $X=\infty$. The solutions at infinity are power series in $1/X$ beginning with a $(1/X)^2$ term. This is the same behaviour as nead $X=0,1$ once we account for the prefactor $1/x_{41}^2x_{23}^2$ that gives the D$_2$-function its weight. At the remaining singularity $X=Z$,  one solution approaches a constant while the other behaves as $1/X^2$.

The singularities when $X=0,1,\infty$ are the standard CFT singularities when boundary points are chosen to collide. The `unexpected' singularity at $X=Z$ also appears in the 4-pt function of full string theory on AdS$_3\times$S$^3$~\cite{Maldacena:2001km}, where it was shown to be associated to a worldsheet instanton that was possible only when the worldsheet and boundary cross-ratios coincide. Exactly these same worldsheet instantons play a key role in the recent work of Eberhardt, Gaberdiel and Gopakumar~\cite{Eberhardt:2019ywk} on the tensionless limit of the AdS$_3$ string, {\it i.e.} the {\it opposite} limit to the supergravity limit. It is fascinating to see evidence of their occurrence also in ambitwistor strings.

However, as we see from~\eqref{eqn:gravity-amplitude}-\eqref{eqn:gauge-amplitude}, the ambitwistor string amplitude formul\ae\ involve the eigenfunctions only on the support of the scattering equations. In the $n=4$ case, this is when $\tau(Z)$ vanishes, so
\begin{equation}
\label{eqn:n=4-scattering}
	\tau(Z)=0\qquad\implies\qquad Z = Z_*\equiv -\frac{\tau_{12}}{\tau_{14}}\,.
\end{equation}
This freezes the integral over the worldsheet moduli space, and may be compared to the flat-space solution $Z = -s/u$ in terms of the usual Mandelstam variables.
 When~\eqref{eqn:n=4-scattering} holds, \eqref{eqn:Heun}  simplifies dramatically and the general solution becomes
\begin{equation}
\label{eqn:eigenfns-when-scatt-eq-hold}
\phi_{\tau}({\bf z}_*;{\bf x}) = \frac{1}{(X-Z_*)^2}
\left[ a + b\,\ln(X^{Z_*}(1-X)^{1-Z_*})\right]
\end{equation}
with $a$ and $b$ constant. These may be used in the amplitude formul\ae. 

\medskip

Unfortunately, obtaining the expansion coefficients the worldsheet correlator in this basis, such as~\eqref{eqn:worldsheet-correlator-expansion}, requires further information which we leave for future work. An important check that such an expansion is indeed possible comes from examining the behaviour of the eigenfunctions in the limit that the worldsheet factorizes. Consider for example the $Z\to0$ factorization channel. In this limit, solutions of \eqref{eqn:Heun} reduce to the hypergeometric function
\begin{equation}
	\left. \phi_\tau({\bf z};{\bf x})\right|_{Z\to 0} \,=\,
	X^{-s} ~	_2F_1(2-s,2-s;2(2-s);X)\,,
\end{equation}
where $s$ solves $\tau_{12} = (s-2)(s-1)$. This is exactly the hypergeometric function that appears in the Mellin representation of the D$_2$-function when the integral over the Mellin $t$ parameter is performed: 
\begin{equation}
\label{eqn:Inverse-Mellin}
\begin{aligned}
	{\rm D}_\Delta(X) &= \int \,\Gamma(s)^2\,\Gamma(t)^2\,
	\Gamma(\Delta-s-t)^2\,X^{-s}(1-X)^{-t}~\diff s\,\diff t \\
	&=\int  \frac{\Gamma(\Delta-s)^4 \,\Gamma(s)^2}{\Gamma(2(\Delta-s))}\, X^{-s} ~ 
	_2F_1(\Delta-s,\Delta-s;2(\Delta-s);X)~\diff s\,.
\end{aligned}
\end{equation}
Similarly, in the $Z\to1$ factorization channel we find \eqref{eqn:Heun} has solution
\begin{equation}
	\left. \phi_\tau({\bf z};{\bf x})\right|_{Z\to 1} \,=\,
	(1-X)^{-t} ~	_2F_1(2-t,2-t;2(2-t);1-X)\,,
\end{equation}
where $t$ solves $\tau_{13} = (t-2)(t-1)$. This is the result of first performing the inverse Mellin transform~\eqref{eqn:Inverse-Mellin} over $s$. This suggests that, although the D-function contains no information  about the worldsheet, it can nonetheless be expanded in a ${\bf z}$-dependent basis. Furthermore, at least in this $n=4$ case, we see a close relation between the Mellin parameters $\delta_{ij}$ and the Gaudin parameters $\tau_{ij}$.

\section{Conclusions}
\label{sec:conclusions}

This paper has begun the investigation of ambitwistor strings on AdS$_3\times$S$^3\times M$.  We verified that the type II model is anomaly free and describes type II supergravity on this background. The worldsheet path integral may be computed explicitly, even at $n$-points, in the case of like-polarization gravitons. We conjectured a natural extension to arbitrary choices of polarization.  The worldsheet punctures are localized to solutions of scattering equations, which is the statement that the spectral parameters are chosen so that all eigenvalues of the $\frak{sl}_2$ Gaudin Hamiltonian vanish. We argued that these AdS scattering equations take essentially the same form as in flat space. 

\medskip

There are many questions that the present work leaves open. The most urgent is to obtain the basis of $n$-particle Gaudin eigenstates and expand the worldsheet correlation functions in this basis. These are the remaining missing steps required to have concrete formul\ae\ for $n$-particle supergravity tree amplitudes in AdS$_3$. Obtaining the eigenfunctions themselves is not expected to be difficult. Indeed, a general framework for doing so was provided long ago by Sklyanin~\cite{Sklyanin:1987ih,Sklyanin:1995bm} as a sophisticated form of separation of variables. To apply this to our case requires, first, understanding the additional constraint that our eigenstates must be invariant under global $SL(2)$ transformations and, second, the right analytic properties and inner product that should be placed on these eigenstates. It seems likely that constraints such as having the correct behaviour under worldsheet factorization,  the boundary OPE, and ensuring the correlator is a single-valued function of the boundary points will play a key role here.

It would also be very  interesting to check that the ambitwistor expressions we have obtained possess the correct flat space limit. This could perhaps be implemented by a version of the Mellin space procedure~\cite{Penedones:2010ue,Fitzpatrick:2011hu}.

While the spectrum includes all states of $d=10$ supergravity on AdS$_3\times$S$^3\times M$, in this paper we have only investigated external states that are independent of the compact manifolds. Turning  on the Kaluza-Klein modes should not lead to any significant difficulties, at least in the case $M=T^4$. In particular, we expect that states that are independent of $M$ may be described by an ${\frak sl}_2\times{\frak sl}_2$ Gaudin 
model, again in representations that have vanishing total quadratic Casimir. Including the S$^3$ involves only finite dimensional representations that may be handled by the algebraic Bethe ansatz.

It would also clearly be interesting to extend this work to other supergroups or cosets with vanishing dual Coxeter number (see also~\cite{Eberhardt:2020draft}). One obvious choice is to take ${\cal G}=PSL(2|2)$, giving an ambitwistor string on the supergroup corresponding to AdS$_3\times S^3$. This could perhaps be done using an ambitwistor version of Berkovits' hybrid string~\cite{Berkovits:1999im}, possibly allowing us to move away from the NS point of the moduli space. More ambitiously, one could consider the $PSU(2,2|4)/(SO(1,4)\times SO(5))$ supercoset. In supergravity, we expect such backgrounds to require RR flux and so call for pure spinors. A pure spinor version of the flat space ambitwistor string was given in~\cite{Berkovits:2013xba},  with a proposal for the extension to AdS$_5\times$S$^5$ given in~\cite{Chandia:2015sfa} at the classical level. However, the origin of the scattering equations remains somewhat mysterious in these models, even in flat space. 

The connection between the ambitwistor string and quantum integrable systems is intriguing and certainly deserves further exploration. Precisely the same Gaudin model arises in a standard WZW model by taking the Knizhnik-Zamolodchikov equation to the critical level $k = -h^\vee$. For general groups ${\cal G}$, WZW correlators at the critical level were computed in~\cite{Feigin:1994in}, where they were shown to be governed by opers on the Langlands dual $^L{\cal G}$. Taking the level to be critical is analogous to sending $\alpha'\to\infty$ in flat space, so this connection is the AdS$_3$ version of the fact that the flat space ambitwistor string localizes on the same points in ${\cal M}_{0,n}$ as appear in the Gross-Mende limit for high-energy, fixed angle scattering. It is intriguing that this relation, which remains mysterious even in flat space, should endure to AdS.

The Gaudin model is well-known to be a linearisation of the XXX spin chain, which is governed by the Yangian ${\cal Y}(\mathfrak{sl}_2)$. In our context, it would be fascinating to understand whether this relation corresponds to deforming the ambitwistor string into the  full string. Conversely, it would be an integrable sector of the boundary CFT, in which the integrable model is the Gaudin model, can help identify the CFT$_2$ dual of pure supergravity. Separation of variables in the XXX spin chain, for exactly the representations arising here, has been studied in~\cite{Faddeev:1994zg,Derkachov:2001yn,Derkachov:2002wz}. Finally, it is natural to wonder whether the elliptic Gaudin model~\cite{Gaudin:1976sv} similarly arises when computing 1-loop ambitwistor string amplitudes.  We note this model is also amenable to separation of variables~\cite{Sklyanin:1998ztv}.

\acknowledgments

We thank Nathan Berkovits, Roland Bittleston, Nick Dorey, David Tong and Beno\^{i}t Vicedo for helpful discussions. We also thank the organisers of the workshop `New Directions in Conventional and Ambitwistor String Theories' at Perimeter Institute, where this work was initiated. The work of DS is supported in part by STFC consolidated grant ST/P000681/1.

\begin{appendix}

\section{Some properties of group manifolds}
\label{app:group-manifolds}

In this appendix we collect some standard results about the geometry of group manifolds that are useful in the text. 
 
\medskip

Given some coordinate chart $x^\mu$ on the group manifold, such that $g = g(x)$, we can expand
 \begin{equation}
 g^{-1} \partial_\mu g \, =  \, e_\mu^a(x)  \, t_a
 \end{equation}
where $e_a^\mu (x)$ is called the left-invariant frame or Maurer-Cartan frame. We can use this to expand the right action on a function $\varphi(x)$ in terms of derivatives. If the generator of right-translations is $j_a$, we have
 \begin{equation}
 \{ j_a , \varphi(x) \}  \, = \,  e_a^\mu \, \, \partial_\mu \varphi (x) \, = \,  e_a \varphi (x) ~.
 \end{equation}
 With this definition it is clear that the vector fields $e_a \equiv e_a^\mu \, \partial_\mu $ satisfy the algebra
 \begin{equation}
 [ e_a , \, e_b ] = \f abc \, e_c 
 \end{equation}
reflecting the group structure.

At various points in the text, we need to extend the action of right-translations to tensors on $G$, rather than simply functions. The simplest way to do this is via the Lie derivative $\mathcal{L}_a$ along $e_a$. We can find the expression for the Lie derivative in the Cartan frame by considering for instance the action on a one-form $V_\mu \, \diff x^\mu \equiv V_a \, (g^{-1} \diff g)^a$.  Right-translations act on this one-form as
\begin{equation}
  t_a^{(R)} \left( \,  V_b (g) \, (g^{-1} \diff g)^b \,  \right)  ~=~ \left( e_a^\mu  \, \partial_\mu V_b  - \f abc \, V_c \right) (g^{-1} \diff g)^b   ~,
 \end{equation}
 so the components $V_a $ transform as $\mathcal{L}_a  \, V_b  = e_a(V_b) - \f abc \, V_c $. As usual, this extends by linearity to tensor fields of higher rank.
 
\medskip

Since our group also comes with a bi-invariant metric $\mathrm{m}$,  instead of the Lie derivative we can also consider right-translating tensor fields using the Levi-Civita connection of $\mathrm{m}$. We now review a few basic properties of the metric $\mathrm{m}$, before considering its Levi-Civita connection and comparing this to the Lie derivative.

We introduce a metric on $G$ by
 \begin{equation}
  \diff s^2 =    \mathrm{m} ( g^{-1} \partial_\mu g , \, g^{-1} \partial_\nu g )  ~ \diff x^\mu \, \diff x^\nu =  \mathrm{m}_{ab} \, e_\mu^a \, e_\nu^b  ~ \diff x^\mu \, \diff x^\nu  ~.
 \end{equation}
on the patch covered by our coordinates $x^\mu$. The inverse/dual frame field $e_\mu^a$ thus serves as vielbein for this metric. Using the group structure, we can always choose the components $\mathrm{m}_{ab}$ in the Maurer-Cartan frame to be constant and right-invariant
 \begin{equation}\label{eqn:metric-invariance}
 \partial_\mu \, \mathrm{m}_{bc} = 0 ~, \qquad \mathrm{m}_{ad} \,  \f bcd  =   \f abd  \,  \mathrm{m}_{cd}    ~.
 \end{equation}
One example for this is the Killing form $\kappa_{ab} \equiv \mathrm{f}_{ac}^d \mathrm{f}_{bd}^c$, but we emphasize that for semi-simple groups there may be several invariant quadratic forms. 

We let $\nabla_\mu$ be the Levi-Civita connection associated to $\mathrm{m}$. Then $\nabla_\mu$, or equivalently $\nabla_a \equiv e_a^\mu \, \nabla_\mu$ in the Maurer-Cartan frame, acts in the usual fashion on tensor fields on the group manifold. For the connection to act on tensors carrying algebra indices, we need the spin connection $\omega_{\mu \, a}^b$, for example on the metric $\mathrm{m}$:
 \begin{equation}\nonumber
 \nabla _\mu \, \mathrm{m}_{ab} =  \partial_\mu \mathrm{m}_{ab} - \omega_{\mu \, a } ^ c \, \mathrm{m}_{cb} - \omega_{\mu \, b } ^ c \, \mathrm{m}_{ac}
 \end{equation}
 and similarly for tensors of other rank. The spin connection is defined by the equation
 \begin{equation}
 \nabla_\mu  e_\nu^a \equiv \partial_\mu  e_\nu^a - \Gamma_{\mu\nu}^\rho \, e_\rho^a + \omega_{\mu \, b}^a \, e_\nu^b  =  0
 \end{equation}
 which can be solved for $\omega$ as
 \begin{equation}\label{eqn:spinconnection-structureconstants}
  \omega_{\mu \, a}^b  =  - \frac{1}{2}   \, \f acb \, e_\mu ^c ~.
 \end{equation}
 This relation between the spin connection and the structure constants means the invariance conditions \eqref{eqn:metric-invariance} and $\nabla_{\! a}  \, \mathrm{m}_{bc} = 0 $ are compatible. Furthermore, it establishes the Riemann tensor in terms of the structure constants
 \begin{equation}
 R_{abcd} = \frac{1}{2} \mathrm{f}_{ab}^d\,  \mathrm{f}_{dc}^e ~.
 \end{equation}
 
 We will need these tools in section~\ref{sec:vertex} to show that BRST cohomology of our model consists of solutions to the supergravity equations, linearized around AdS$_3\times S^3$.

\end{appendix}


\bibliographystyle{JHEP}
\bibliography{bibliography}

\end{document}